\documentclass[conference,compsoc]{IEEEtran}

\PassOptionsToPackage{table}{xcolor}
\usepackage[nocompress]{cite}
\usepackage{url}
\usepackage{enumitem}
\usepackage{array}
\usepackage{booktabs}
\usepackage{xspace}
\usepackage{tikz}
\usepackage{tabularx}
\usepackage{xcolor}
\usepackage{microtype}
\usepackage{placeins}
\usepackage{needspace}
\usepackage[hidelinks,breaklinks=true]{hyperref}
\usepackage{amssymb}
\usepackage{tcolorbox}
\tcbuselibrary{skins,breakable}
\usepackage{ragged2e}
\newtcolorbox{textbox}{
    colback=gray!5,
    colframe=gray!75,
    boxrule=0.5pt,
    left=6pt,
    right=6pt,
    top=6pt,
    bottom=6pt,
    breakable,
    width=\linewidth,
    before upper={\sloppy}
}

\microtypesetup{expansion=false}
\usetikzlibrary{positioning,shapes.geometric,arrows.meta,calc,backgrounds,decorations.markings}

\definecolor{privRed}{HTML}{B2182B}
\definecolor{privBg}{HTML}{FDECEA}
\definecolor{pubBlue}{HTML}{4477AA}
\definecolor{pubBg}{HTML}{EEF3FA}
\definecolor{teal}{HTML}{228833}
\definecolor{tealBg}{HTML}{E6F4F1}
\definecolor{neutralGray}{HTML}{4A5568}
\definecolor{midgray}{HTML}{666666}
\definecolor{stagefill}{HTML}{F5F5F5}
\definecolor{darkgray}{HTML}{3D4349}
\definecolor{warnBorder}{HTML}{C0490A}
\definecolor{warnBg}{HTML}{FEF3EB}
\definecolor{warnText}{HTML}{7D3000}
\definecolor{passGreen}{HTML}{228833}
\definecolor{passBg}{HTML}{EBF7EF}
\definecolor{armOrange}{HTML}{9A3A00}
\definecolor{armOrangeBg}{HTML}{FEF0E6}
\definecolor{axisGray}{HTML}{667085}
\definecolor{linkGray}{HTML}{8A94A6}
\definecolor{dashgray}{HTML}{9BA4AF}
\definecolor{armZero}{HTML}{1A5276}
\definecolor{armZeroBg}{HTML}{E8F0FB}
\definecolor{armPolicy}{HTML}{117864}
\definecolor{armPolicyBg}{HTML}{E6F4F1}
\definecolor{humanGray}{HTML}{6F7782}
\definecolor{humanBg}{HTML}{F2F4F7}
\definecolor{execblue}{HTML}{4477AA}

\newcolumntype{L}{>{\raggedright\arraybackslash}X}
\newcolumntype{C}{>{\centering\arraybackslash}X}

\newcommand{\enterprisetotalattackpatterncount}{835}
\newcommand{\enterprisedeprecatedattackpatterncount}{144}
\newcommand{\enterprisetotalintrusionsetcount}{187}
\newcommand{\enterprisetotalmalwarecount}{696}
\newcommand{\enterprisetotaltoolcount}{91}
\newcommand{\enterprisetotalrelationshipcount}{20048}
\newcommand{\enterprisetacticcount}{14}
\newcommand{\enterpriseplatformcount}{691}
\newcommand{\enterpriseplatformpct}{100.0}
\newcommand{\enterprisesystemrequirementscount}{0}
\newcommand{\enterprisesystemrequirementspct}{0.0}
\newcommand{\mobileplatformpct}{100.0}
\newcommand{\icsplatformpercentage}{98.8}
\newcommand{\fightplatformpercentage}{100.0}
\newcommand{\enterprisecampaignswithsoftwarecount}{47}
\newcommand{\enterprisecampaignswithsoftwarepercentage}{90.4}
\newcommand{\enterpriseactivecampaigncount}{52}
\newcommand{\enterprisecampaignswithplatformsignalcount}{47}
\newcommand{\enterprisecampaignswithplatformsignalpct}{90.4}
\newcommand{\enterprisecampaignsplatformunknowncount}{5}
\newcommand{\enterprisecampaignsplatformunknownpct}{9.6}
\newcommand{\campaignoswindowscount}{42}
\newcommand{\campaignoslinuxcount}{23}
\newcommand{\campaignosmacoscount}{21}
\newcommand{\campaignoswindowspctsignal}{89.4}
\newcommand{\campaignoslinuxpctsignal}{48.9}
\newcommand{\campaignosmacospctsignal}{44.7}
\newcommand{\campaignoswindowspcttotal}{80.8}
\newcommand{\campaignoslinuxpcttotal}{44.2}
\newcommand{\campaignosmacospcttotal}{40.4}
\newcommand{\campaignsuttiercoarsecount}{47}
\newcommand{\enterpriseintrusionsetswithsoftwarecount}{158}
\newcommand{\enterpriseintrusionsetswithsoftwarepercentage}{91.9}
\newcommand{\enterpriseactiveintrusionsetcount}{172}
\newcommand{\enterpriseactivesoftwarecount}{784}
\newcommand{\enterpriseactivemalwarecount}{693}
\newcommand{\enterpriseactivetoolcount}{91}
\newcommand{\softwarewithversionsignalpercentage}{2.4}
\newcommand{\softwarewithcpepercentage}{0.0}

\newcommand{\softwareversionenrichmentgaincount}{6}
\newcommand{\softwareversionenrichedtotalpct}{3.2}
\newcommand{\softwareversionenrichmentgainpp}{0.8}
\newcommand{\cveuniquecount}{26}
\newcommand{\cvefreetextonlycount}{5}
\newcommand{\cvefromfreetextpct}{19.2}
\newcommand{\cveactionablecount}{12}
\newcommand{\cvetechniqueonlycount}{14}
\newcommand{\campaignlinkedcvecount}{8}
\newcommand{\entcampaignswithcvestructuredcount}{2}
\newcommand{\entcampaignswithcvestructuredpct}{3.8}
\newcommand{\entcampaignswithcveenrichmentgaincount}{3}
\newcommand{\entcampaignswithcveenrichmentgainpp}{5.8}
\newcommand{\entcampaignswithcvecount}{5}
\newcommand{\entcampaignswithcvepct}{9.6}
\newcommand{\entintrusionsetswithcvestructuredcount}{2}
\newcommand{\entintrusionsetswithcvestructuredpct}{1.2}
\newcommand{\entintrusionsetswithcveenrichmentgaincount}{2}
\newcommand{\entintrusionsetswithcveenrichmentgainpp}{1.1}
\newcommand{\entintrusionsetswithcvecount}{4}
\newcommand{\entintrusionsetswithcvepct}{2.3}
\newcommand{\campaignswithinitialaccesscount}{38}
\newcommand{\campaignswithinitialaccesspct}{73.1}
\newcommand{\campaignswithinitialaccessnocvecount}{33}
\newcommand{\campaignswithinitialaccessnocvepct}{63.5}
\newcommand{\compatibilitycontainerfeasiblecount}{19}
\newcommand{\compatibilitycontainerfeasiblepercentage}{2.7}
\newcommand{\compatibilityvmrequiredcount}{526}
\newcommand{\compatibilityvmrequiredpercentage}{76.1}
\newcommand{\compatibilityinfrastructuredependentcount}{146}
\newcommand{\compatibilityinfrastructuredependentpercentage}{21.1}
\newcommand{\compatibilityexplicitassignmentpercentage}{32.7}
\newcommand{\compatibilityfallbackassignmentpercentage}{67.3}
\newcommand{\compatibilitynoncffloorpercentage}{29.9}
\newcommand{\compatibilitynoncfceilingpercentage}{97.2}
\newcommand{\compatibilityrulecoveragepercentage}{32.7}
\newcommand{\compatibilitynoncfresolvedpercentage}{91.6}
\newcommand{\compatibilityresolutiongainpp}{58.9}
\newcommand{\compatibilityvalidationsamplesize}{36}
\newcommand{\sutprofileuniquesoftwarepercentage}{90.7}
\newcommand{\sutprofileuniquesoftwarecvepercentage}{90.7}
\newcommand{\sutprofileconfusionsoftwarepercentage}{9.3}
\newcommand{\sutprofileconfusionsoftwarecvepercentage}{9.3}
\newcommand{\thresholdkoneconfusionpct}{1.3}
\newcommand{\thresholdktwoconfusionpct}{0.0}
\newcommand{\thresholdkonesample}{158}
\newcommand{\thresholdktwosample}{132}
\newcommand{\thresholdkthreesample}{105}
\newcommand{\thresholdkfivesample}{76}
\newcommand{\deltazerozerofiveconfusionpct}{9.3}
\newcommand{\bootstrapconfusioncilow}{5.2}
\newcommand{\bootstrapconfusioncihigh}{14.0}
\newcommand{\bootstrapuniquecilow}{86.0}
\newcommand{\bootstrapuniquecihigh}{94.8}
\newcommand{\nullmodeliterations}{1000}
\newcommand{\nullmodelconfusionplowpct}{8.1}
\newcommand{\nullmodelconfusionphighpct}{10.5}
\newcommand{\nullmodelpvalue}{0.34}
\newcommand{\campaignmeantechniquecount}{19.6}
\newcommand{\campaignmediantechniquecount}{17}
\newcommand{\campaignmeantacticcoverage}{8.0}
\newcommand{\campaigncompletekillchaincount}{44}
\newcommand{\campaigncompletekillchainpct}{84.6}
\newcommand{\ieircount}{2}
\newcommand{\ieirpct}{3.8}
\newcommand{\campaignswithosfamilysignalcount}{44}
\newcommand{\campaignswithsingleosfamilycount}{22}
\newcommand{\campaignswithdualosfamilycount}{2}
\newcommand{\campaignswithtripleosfamilycount}{20}
\newcommand{\platformablationtechniquesingleoscount}{1}
\newcommand{\platformablationtechniquetripleoscount}{51}
\newcommand{\onesoftwarecampaignswithossignalcount}{17}
\newcommand{\onesoftwarecampaignssingleoscount}{10}
\newcommand{\onesoftwarecampaignstripleoscount}{7}
\newcommand{\campaignswithnonosonlyplatformsignalcount}{3}
\newcommand{\evidenceconvergenceratepct}{96.2}

\newcommand{\provenanceprofilecount}{20}
\newcommand{\provenancetotalelementcount}{139}
\newcommand{\provenancecorpussupportedcount}{34}
\newcommand{\provenancecorpussupportedpct}{24.5}
\newcommand{\provenanceautosutconcretizedcount}{23}
\newcommand{\provenanceautosutconcretizedpct}{16.5}
\newcommand{\provenanceanalystauthoredcount}{82}
\newcommand{\provenanceanalystauthoredpct}{59.0}
\newcommand{\provenanceplatformtotal}{27}
\newcommand{\provenanceplatformcorpus}{27}
\newcommand{\provenancetopologytotal}{3}
\newcommand{\provenancetopologyanalyst}{3}
\newcommand{\provenanceartifacttotal}{32}
\newcommand{\provenanceartifactanalyst}{32}
\newcommand{\provenancesoftwaretotal}{23}
\newcommand{\provenancesoftwareanalyst}{22}
\newcommand{\provenancevulnerabilitytotal}{3}
\newcommand{\provenancevulnerabilitycorpus}{2}

\newif\ifartifactanonymous
\artifactanonymoustrue

\providecommand{\stix}{STIX\xspace}
\providecommand{\mitre}{MITRE\xspace}
\providecommand{\attack}{ATT\&CK\xspace}
\providecommand{\caldera}{Caldera\xspace}

\providecommand{\Description}[1]{}

\newcommand{\myparagraph}[1]{\emph{#1.}\nobreak\hspace{0.35em}}

\begin{document}

\title{AutoSUT: The Environment Semantics Gap in Structured CTI \\for Adversary Emulation}

%\author{Blinded for review}

%\iffalse
\author{
    \IEEEauthorblockN{Sidnei Barbieri, Ágney Lopes Roth Ferraz, and Lourenço Alves Pereira Júnior}
    \IEEEauthorblockA{
        Computer Science Division, Aeronautics Institute of Technology (ITA)\\
        São José dos Campos, SP, Brazil\\
        {sidneisb@ita.br, roth@ita.br, ljr@ita.br}
    }
}
%\fi

\maketitle

\begin{abstract}
Structured Cyber Threat Intelligence (CTI) increasingly supports adversary emulation, detection evaluation, and cyber range design, yet each workflow still requires a target System Under Test (SUT) whose environment is not fully described by public CTI. We define the environment semantics gap as a measurable property of structured CTI: the SUT information required for replay-ready instantiation that cannot be derived solely from structured fields. We present AutoSUT, a pipeline that locates where corpus-supported narrowing ends and analyst specification begins. Across ATT\&CK Enterprise, Mobile, and ICS STIX bundles, with CAPEC and FiGHT as contrast datasets, we measure platform coverage, software specificity, vulnerability evidence, and deployment compatibility. Platform tags are near-universal, but 97.6\% of Enterprise software objects lack version indicators and CPE identifiers. Campaign-level CVE evidence covers only 9.6\% of campaigns, even after free-text enrichment, and only 19 of 691 techniques (2.7\%) are container-feasible under conservative backend-family assignment.

Profile confusion among intrusion sets drops from 1.3\% for one linked software item to 0\% for two linked software items, indicating that software-evidence density, not CVE enrichment, drives actor-specific SUT screening. Finally, we constructively demonstrate environment non-uniqueness: holding every corpus-supported element fixed and varying only the analyst-authored region yields multiple distinct, campaign-compatible SUTs, including an executable witness running CVE-2021-41773 and coincident witnesses in which structurally different service realizations execute the same attack. Structured CTI, therefore, constrains but does not uniquely determine the executable environment. Replay-ready emulation should accordingly declare which environment commitments the corpus supports and which remain analyst-authored.
\end{abstract}

\begin{IEEEkeywords}
cyber threat intelligence, STIX, MITRE ATT\&CK, system under test, adversary emulation, structured CTI
\end{IEEEkeywords}

%===============================================================
\section{Introduction}\label{sec:introduction}

Structured Cyber Threat Intelligence (CTI) is increasingly used to support adversary emulation, detection evaluation, cyber range construction, and attribution analysis. The execution-facing workflows among them need two inputs: a description of \emph{what} an adversary does and an environment in which those actions can run. The first is increasingly well served by the \mitre{} \attack{} corpus, which catalogs \enterprisetotalattackpatterncount{} Enterprise techniques across \enterprisetotalintrusionsetcount{} documented groups and \enterpriseactivecampaigncount{} campaigns, all encoded in machine-readable Structured Threat Information Expression (\stix{}) bundles~\cite{strom2018mitre,DBLP:conf/uss/VirkudIR00024}. The second remains structurally underspecified. Each replay or evaluation use still needs a \emph{System Under Test} (SUT): a target environment with the intended operating system, software stack, exposure surface, and vulnerability state. If that environment is wrong, even a correct procedure cannot produce a reproducible emulation result.

We therefore define the \emph{environment semantics gap} \footnote{\url{https://anonymous.4open.science/r/autosut-reproducibility-artifact-3EBE/} - The artifact will be publicly released after the completion of the double-blind review process.} as a measurable property of structured CTI: the SUT information required to instantiate a replay-ready environment that cannot be derived from structured CTI alone. We ask: \emph{what is the strongest lower-bound SUT claim that public ATT\&CK-style structured CTI can justify without external enrichment?} The input is public structured CTI in ATT\&CK-style \stix{} bundles. The output is a bounded SUT claim: which environment family the corpus narrows, which backend family it justifies, and which layers remain analyst-authored. AutoSUT is the measurement pipeline: it locates where corpus-supported narrowing ends and analyst specification begins, without substituting exploit synthesis, command generation, or runtime inference for missing CTI. We then add one controlled construction, a non-uniqueness demonstration (Section~\ref{sec:nonuniqueness}), that varies only the analyst-authored region to show the measured gap has an executable consequence.

Across the \attack{} Enterprise, Mobile, and Industrial Control Systems (ICS) datasets, with the Common Attack Pattern Enumeration and Classification (CAPEC) and the 5G Hierarchy of Threats (FiGHT) used as coverage contrasts, the pattern is consistent. Platform evidence is common but coarse. Software references rarely pin versions or Common Platform Enumeration (CPE) identifiers. Common Vulnerabilities and Exposures (CVE) evidence is sparse and often split between structured fields and free text. The result is a reproducible quantitative boundary between what public datasets encode and what analysts must still supply.

That boundary matters to every downstream consumer of structured CTI. Red-team platforms execute only after the environment has been declared~\cite{applebaum2016caldera,redcanary2023art,orbinato2024laccolith,damodaran2025automated,wang2024sands}; MITRE Engenuity ATT\&CK evaluation programs still depend on manually declared representative environments~\cite{10.1145/3634737.3645012}; provenance-graph detection systems build labs whose representativeness depends on how completely CTI specifies the target environment~\cite{milajerdi2019holmes,han2020unicorn,DBLP:conf/uss/Dong0NS0LLX23,DBLP:conf/uss/YangXXLZ23}; and attribution confidence is bounded by evidence quality rather than by the absence of behavioral hypotheses~\cite{DBLP:conf/uss/SahaMBCVL25,DBLP:conf/ccs/YuldoshkhujaevJ25}. An under-specified environment context can therefore distort detection evaluation, emulation fidelity, and attribution claims alike~\cite{DBLP:conf/ccs/LvGQCZCJ24}. Many execution-facing workflows implicitly rely on structured CTI to narrow the environment in which documented behavior runs; how much of that environment the corpus actually determines is not directly measured.

ATT\&CK is an appropriate public corpus in which to test this boundary because it is the most widely reused structured CTI source for threat-informed defense and because its public STIX bundles expose behavior-centric evidence in a machine-readable form~\cite{strom2018mitre,DBLP:conf/uss/VirkudIR00024,10.1145/3634737.3645012,DBLP:conf/ccs/DongLJLWHXCLGC23}. The question is not whether ATT\&CK or STIX should be a lab generator, but how far their public structured fields already determine the environment before analyst authorship resumes. We therefore isolate the environment dimension from the procedural one. This leaves four research questions (RQs):

\noindent\hangindent=2.8em\hangafter=1
\textbf{RQ1:} How much structured CTI encodes the platform, software, and vulnerability details needed for SUT construction?

\noindent\hangindent=2.8em\hangafter=1
\textbf{RQ2:} How far can those signals support lower-bound SUT and backend-family claims, and where do they stop?

\noindent\hangindent=2.8em\hangafter=1
\textbf{RQ3:} How specific are SUT requirement profiles for campaigns and intrusion sets without over-claiming attribution?

\noindent\hangindent=2.8em\hangafter=1
\textbf{RQ4:} Given the corpus-supported elements a SUT must preserve, does structured CTI determine a unique environment, or admit multiple compatible ones?

This paper provides the first measurement of \emph{environment derivability} from ATT\&CK-style structured CTI, namely how much of a replay-ready SUT the public corpus determines, as distinct from the behavioral and field-completeness quality that prior studies assess. The contributions are fourfold. We define a SUT-centric measurement model for the environment semantics gap; measure where platform, software, vulnerability, and backend signals remain strong enough for coarse narrowing and where they stop; quantify bounded enrichment, including a narrow CVE-to-package concretization slice; and constructively demonstrate \emph{environment non-uniqueness} with executable witnesses: one running the disclosed CVE-2021-41773 and two in which structurally different service realizations run the same attack (Section~\ref{sec:nonuniqueness}). Taken together, these results support a single thesis: structured CTI fixes an invariant environmental core but leaves a free region, so it constrains but does not uniquely determine the executable environment, and that under-determination has a demonstrable, executable consequence.

\begin{figure}[htpb]
\centering
\makebox[\columnwidth][l]{\hspace*{-0.05\columnwidth}\resizebox{1.05\columnwidth}{!}{\providecolor{execblue}{HTML}{4477AA}
\providecolor{loopgreen}{HTML}{228833}
\providecolor{ctired}{HTML}{EE6677}
\providecolor{edgegray}{HTML}{333333}
\providecolor{midgray}{HTML}{666666}
\providecolor{dashgray}{HTML}{BBBBBB}
\providecolor{stagefill}{HTML}{F5F5F5}

\begin{tikzpicture}[
  x=1cm,y=1cm,
  font=\sffamily\footnotesize,
  >=latex,
  line cap=round,
  line join=round
]
  \path[use as bounding box] (-0.32,-1.48) rectangle (8.32,6.78);

  \fill[execblue!2,rounded corners=5pt] (1.20,6.40) rectangle (8.35,0.30);
  \draw[execblue!20,rounded corners=5pt,line width=0.45pt] (1.20,6.40) rectangle (8.35,0.30);

  \node[draw=execblue,rounded corners=3pt,fill=execblue!8,text=edgegray,
        minimum width=6.85cm,minimum height=1.04cm,align=center,line width=0.8pt,
        text width=6.65cm,inner xsep=4pt,inner ysep=2pt] (p) at (4.78,5.75)
    {\textbf{Platform tags} $\rightarrow$ OS / device / cloud family};

  \node[draw=loopgreen,rounded corners=3pt,fill=loopgreen!8,text=edgegray,
        minimum width=5.95cm,minimum height=1.04cm,align=center,line width=0.8pt,
        text width=5.65cm,inner xsep=4pt,inner ysep=2pt] (s) at (4.78,4.15)
    {\textbf{Software links} $\rightarrow$ installable product\\family / compatibility envelope};

  \node[draw=ctired,rounded corners=3pt,fill=ctired!8,text=edgegray,
        minimum width=5.05cm,minimum height=1.04cm,align=center,line width=0.8pt,
        text width=4.55cm,inner xsep=4pt,inner ysep=2pt] (v) at (4.78,2.55)
    {\textbf{Campaign CVE} $\rightarrow$ plausible\\vulnerable target class};

  \node[draw=edgegray!70,rounded corners=3pt,fill=stagefill,text=edgegray,
        minimum width=4.15cm,minimum height=1.04cm,align=center,line width=0.75pt,
        text width=3.80cm,inner xsep=4pt,inner ysep=2pt] (b) at (4.78,0.95)
    {\textbf{CF / VMR / ID} $\rightarrow$ minimum\\backend family};

  \draw[->,edgegray!80,line width=1.0pt,shorten >=1.5pt,shorten <=1.5pt] (p.south) -- (s.north);
  \draw[->,edgegray!80,line width=1.0pt,shorten >=1.5pt,shorten <=1.5pt] (s.south) -- (v.north);
  \draw[->,edgegray!80,line width=1.0pt,shorten >=1.5pt,shorten <=1.5pt] (v.south) -- (b.north);

  \node[draw=dashgray!90!edgegray,dashed,rounded corners=3.5pt,fill=stagefill,
        minimum width=6.85cm,minimum height=1.20cm,align=center,text=edgegray,
        text width=6.35cm,inner sep=4pt,line width=0.80pt] (r) at (4.78,-0.73)
    {\textbf{Still analyst-supplied}\\exact version, host configuration, trust boundary, topology, and service layout};

  \node[anchor=east,text=midgray,font=\sffamily\scriptsize\bfseries,align=right,text width=1.62cm]
    at (1.03,5.75) {broadest\\candidate\\SUT space};

  \node[anchor=east,text=midgray,font=\sffamily\scriptsize\bfseries,align=right,text width=1.62cm]
    at (1.03,0.95) {strongest\\structured\\lower bound};

  \draw[dashgray!95!edgegray,line width=0.55pt] (1.05,5.85) -- (1.05,0.90);
\end{tikzpicture}}}
\caption{\label{fig:measurement_model}From structured CTI to partial SUT design. Each evidence layer yields only a lower-bound environment claim; the analyst still supplies the exact version, host configuration, and topology needed for replay-ready deployment.}
\Description{Conceptual diagram showing four descending evidence layers for SUT derivation: platform tags narrow OS, device, or cloud family; software links narrow installable product family; campaign CVE narrows the plausible vulnerable target class; and the CF, VMR, and ID taxonomy narrows the minimum backend family. A dashed box below marks the analyst-supplied remainder: exact version, host configuration, trust boundary, topology, and service layout.}
\end{figure}
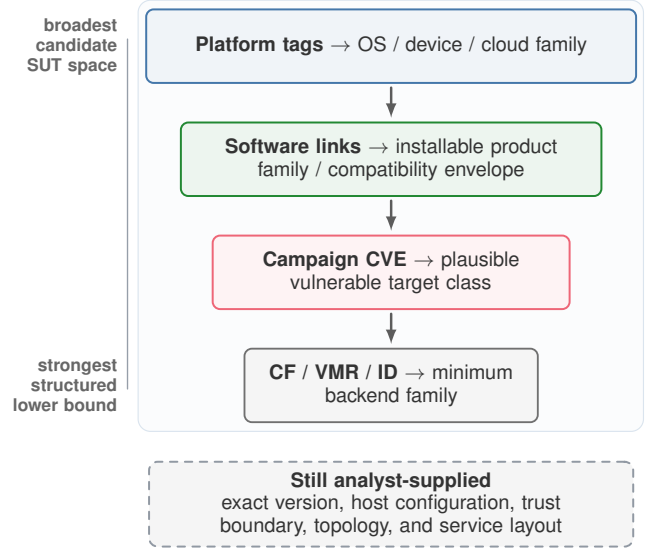

% ===============================================================
\section{Background}\label{sec:background}

The \mitre{} \attack{} (Adversarial Tactics, Techniques, and Common Knowledge) framework documents adversary behavior as tactics, techniques, and sub-techniques across three domains: Enterprise, Mobile, and ICS. Enterprise is the largest of the three, containing \enterprisetotalattackpatterncount{} techniques and sub-techniques across \enterprisetacticcount{} tactics in v18.1~\cite{strom2018mitre,DBLP:conf/uss/VirkudIR00024}. ATT\&CK content is distributed as \stix{} 2.x bundles~\cite{stixspec}. For SUT construction, the most relevant object types are attack patterns, campaigns, intrusion sets, malware, tools, and their relationships (Table~\ref{tab:entities}). In the public ATT\&CK bundles we measure, STIX captures semantic associations such as \emph{uses}, \emph{attributed-to}, and \emph{targets}; these objects do not, by themselves, specify deployment-ready environment constraints at the granularity needed for SUT derivation.

\begin{table}[htpb]
\caption{ATT\&CK entity types and their SUT relevance.}\label{tab:entities}
\centering\small
\setlength{\tabcolsep}{3pt}
{\rowcolors{2}{stagefill}{white}
\begin{tabularx}{\columnwidth}{@{}>{\raggedright\arraybackslash}p{0.22\columnwidth}L@{}}
\toprule
\textbf{Entity} & \textbf{SUT relevance} \\
\midrule
\emph{attack-pattern} &
  Platform hints via \nolinkurl{x_mitre_platforms}; system requirements in \nolinkurl{x_mitre_system_requirements} (rarely populated). \\
\emph{malware} / \emph{tool} &
  Implies OS dependency; version rarely specified; CPE mappings largely absent. \\
\emph{campaign} &
  Scope operation to the time window and target set; links to techniques and intrusion set. \\
\emph{intrusion-set} &
  Aggregates behavioral profile; useful for SUT susceptibility profiling but not for single-campaign instantiation. \\
\emph{vulnerability} &
  CVE-mapped objects are present in some bundles; they are rarely linked directly to campaigns. \\
\bottomrule
\end{tabularx}}
\end{table}

Red-team platforms such as \caldera{} execute adversary procedures against a declared target environment rather than deriving that environment directly from CTI~\cite{applebaum2016caldera}. SUT construction, therefore, remains a manual precondition for emulation, which is exactly the burden this paper measures.

% ===============================================================
\section{SUT Measurement Model}\label{sec:definitions}
An \emph{environment} is the complete execution context required to reproduce behavior described by a CTI object: OS family and version, installed software with configurations, network topology, identity and privilege model, and exploitable vulnerabilities. An environment description is \emph{machine-actionable} for SUT derivation when every constraint appears in a structured, parseable field such as \texttt{x\_mitre\_platforms}, a CPE  string, or a CVE identifier rather than in free-text prose. The  \emph{System Under Test} (SUT) is the concrete instantiation of such an environment in a virtualized or containerized lab.

Each CTI object contributes up to three kinds of SUT constraints.  \emph{Platform constraints} come from \texttt{x\_mitre\_platforms} and indicate an OS family, cloud provider, or device type. \emph{Software constraints} arise when a \texttt{malware} or \texttt{tool} object is linked via \texttt{uses} relationships to a technique or campaign. \emph{Vulnerability constraints} arise when a CVE identifier appears in structured fields or when a \texttt{vulnerability} object is linked to a campaign, technique, or software object.

For each campaign or intrusion set, the \emph{SUT requirement profile} is the union of those three components. It is a static analysis construct over a fixed bundle snapshot: the pipeline emits the strongest lower-bound SUT claim that snapshot justifies, without adapting online during execution. We distinguish three profile tiers: \emph{coarse} (platform plus software), \emph{anchored} (coarse plus version/CPE or campaign-level CVE), and \emph{exploit-pinned} (coarse plus campaign-level CVE).

We additionally classify each technique by the minimum backend it requires: \emph{container-feasible} (CF) for user-space behaviors without kernel, firmware, or Active Directory dependence; \emph{virtual-machine-required} (VMR) for kernel interaction, process injection, or boot-level persistence; 
and \emph{infrastructure-dependent} (ID) for behaviors that presuppose a multi-host topology or enterprise identity services.

Each evidence layer only narrows the candidate SUT or backend family (Figure~\ref{fig:measurement_model}); it never fills the \emph{analyst-authored remainder} with the exact version, topology, trust boundary, and exposure details that are still required for replay-ready instantiation. This paper measures the line between environment semantics that can be clearly retrieved from structured CTI fields and those that depend on unstructured evidence or analyst interpretation.

% ===============================================================
\section{Data Sources}\label{sec:dataset}

Following the measurement model in Section~\ref{sec:definitions}, we select datasets that expose the same core STIX structures while covering different ATT\&CK scopes. The \mitre{} \attack{} Enterprise bundle v18.1 is the primary source for all quantitative analyses because it is the largest domain by object count and carries the campaign coverage that RQ2 and RQ3 require. Table~\ref{tab:dataset} reports the raw object counts across the five sources that this study traverses.

\begin{table}[htpb]
\caption{STIX object totals across ATT\&CK domains and sources. Active counts after deprecation filtering are reported in Section~\ref{sec:methodology}.}\label{tab:dataset}
\centering\small
{\setlength{\tabcolsep}{4pt}
\rowcolors{1}{stagefill}{white}
\begin{tabular}{@{}lrrrrr@{}}
\toprule
\textbf{Source} &
\textbf{Techniques} &
\textbf{Campaigns} &
\textbf{IS} &
\textbf{Malware} &
\textbf{Tools} \\
\midrule
Enterprise & \enterprisetotalattackpatterncount{} & \enterpriseactivecampaigncount{}  & \enterprisetotalintrusionsetcount{} & \enterprisetotalmalwarecount{} & \enterprisetotaltoolcount{} \\
Mobile     & 190 & 3   & 17  & 121 & 2  \\
ICS        & 95  & 8   & 16  & 30  & 0  \\
CAPEC      & 615 & 0   & 0   & 0   & 0 \\
FiGHT      & 707 & 0   & 136 & 475 & 73 \\
\bottomrule
\end{tabular}}
\end{table}

RQ2 and RQ3 remain focused on Enterprise because only Enterprise contains sufficient campaign coverage for deployment-oriented analysis; CAPEC is included as STIX-encoded attack-pattern knowledge for contrast in coverage. All RQ metrics use active-object counts after deprecation and revocation filtering. In Enterprise, this filter leaves \enterpriseplatformcount{} active techniques, \enterpriseactivecampaigncount{} campaigns, \enterpriseactiveintrusionsetcount{} intrusion sets, and \enterpriseactivesoftwarecount{} software objects (\enterpriseactivemalwarecount{} malware~$+$~\enterpriseactivetoolcount{} tools), reachable through \enterprisetotalrelationshipcount{} relationship objects.

% ===============================================================
\section{Methodology}\label{sec:methodology}

The pipeline applies explicit extraction and decision rules. We first parse and normalize the bundles, then define the compatibility taxonomy used for deployability claims, bound the enrichment steps allowed beyond structured fields, and specify the profile-specificity metric used in RQ3. Each reported number, therefore, maps to a stable rule rather than to ad hoc interpretation.

\myparagraph{Evidence extraction} We parse five locally available \stix{} bundles as JavaScript Object Notation (JSON) and normalize object identifiers. Enterprise, Mobile, and ICS drive the core analyses; CAPEC and FiGHT support cross-corpus coverage in RQ1. We exclude deprecated and revoked objects, which removes \enterprisedeprecatedattackpatterncount{} of the \enterprisetotalattackpatterncount{} Enterprise attack-pattern objects, leaving \enterpriseplatformcount{} active techniques. Relationships are indexed by (\textit{source\_ref}, \textit{relationship\_type}, \textit{target\_ref}) triples to enable bidirectional traversal. For each campaign, the set of techniques is collected from all \texttt{attack-pattern} objects reachable via \texttt{uses} relationships; software and CVE associations are derived analogously.

From \texttt{attack-pattern} objects, we extract platform scope, host requirements, privilege requirements, and defense-bypass cues. From malware and tool objects, we extract names, aliases, platform tags, and any CPE or version strings in \nolinkurl{external_references}. CVE identifiers are extracted from \nolinkurl{description} fields and from linked vulnerability objects. For campaign-level target-platform inference, we adopt a strict software-only rule. Platforms are inferred only from software-linked platform tags. Technique-level platforms are excluded from this step because they encode general compatibility rather than campaign-specific target evidence. Campaigns with no linked software are reported as \emph{platform unknown} rather than receiving inferred values. This avoids turning generic compatibility into campaign-specific target claims.

\myparagraph{Compatibility taxonomy} The taxonomy is an operational lower-bound abstraction for backend-family assignment, not a claim about full execution fidelity. Each technique receives exactly one compatibility class under a fixed precedence order: container-feasible (\textbf{CF}), virtual-machine-required (\textbf{VMR}), or infrastructure-dependent (\textbf{ID}). Identity and infrastructure cues dominate, host-semantic VM cues come next, and narrow container-feasible cues come last. Any remaining unresolved technique defaults to \textbf{VMR}.

More concretely, we assign \textbf{ID} when platform tags reference \texttt{Windows Domain} or \texttt{Identity Provider}, or when a lateral-movement technique depends on Active Directory software. We assign \textbf{VMR} when permissions-required fields indicate administrator privileges (Administrator, SYSTEM, or root), and likewise for privilege-escalation and boot/firmware defense-evasion techniques. We assign \textbf{CF} when \texttt{x\_mitre\_platforms} is restricted to \texttt{Containers} or \texttt{Linux} and the technique is not in kernel-interaction or boot-persistence clusters.

This taxonomy is a conservative proxy for deployability rather than an execution ground truth. Explicit rules cover \compatibilityexplicitassignmentpercentage\% of techniques; the remaining \compatibilityfallbackassignmentpercentage\% default to VMR so that unresolved cases overestimate rather than underestimate infrastructure cost. CF is narrow, ID captures explicit enterprise-service dependence, and the unresolved middle goes to VMR because the safer failure mode for this paper is false VMR inflation rather than false CF optimism.

\begin{table}[htpb]
\caption{Compatibility classes as conservative lower-bound backend claims: container-feasible (CF), virtual-machine-required (VMR), and infrastructure-dependent (ID).}\label{tab:compatibility_design}
\centering\small
{\setlength{\tabcolsep}{1.5pt}
\rowcolors{2}{stagefill}{white}
\begin{tabularx}{\columnwidth}{@{}l>{\raggedright\arraybackslash}p{0.428\columnwidth}L@{}}
\toprule
\textbf{Class} & \textbf{Structured trigger} & \textbf{Operational reading and preferred conservative error} \\
\midrule
CF & Platforms restricted to Linux/Containers; no kernel, boot, or identity cues & Candidate for container-only replay. The preferred error is false exclusion from CF rather than false CF optimism. \\
VMR & Privilege, kernel, firmware, or boot semantics; also, the unresolved residue & Requires host-level VM semantics. The preferred error is false VMR inflation when structured fields are too weak to justify CF or ID. \\
ID & Domain, identity-provider, or Active Directory semantics; lateral movement with enterprise-service dependence & Requires enterprise services or multi-host structure. The class bounds backend family, not the exact topology, trust boundary, or host count. \\
\bottomrule
\end{tabularx}}
\end{table}

Under that conservative error budget (Table~\ref{tab:compatibility_design}), if all unresolved techniques were optimistically forced to CF, the non-CF share would still be \compatibilitynoncffloorpercentage\%; under VMR or ID defaults, it rises to \compatibilitynoncfceilingpercentage\%. More importantly, the explicit-rule subset is already \compatibilitynoncfresolvedpercentage\% non-CF. The fallback policy, therefore, changes the absolute split, but not the substantive conclusion that container-feasible techniques are a small minority and that unresolved cases mostly inflate VMR rather than hide CF coverage.

For independent auditing, we draw a stratified manual-validation sample of \compatibilityvalidationsamplesize{} techniques, balanced across CF/VMR/ID predictions and covering both explicit rules and fallback-driven VMR cases, so that the narrow CF minority, the explicit ID evidence, and the conservative middle can be inspected separately. The RQ2 percentages remain deterministic rule outputs, while this sample provides the exact rows needed to inspect the construct directly.

\myparagraph{Bounded enrichment} To quantify the practical impact of the measured gaps, we evaluate three bounded enrichment steps over the ATT\&CK bundles used in this study: description-based regex mining for software versions, free-text extraction of campaign-level CVE references, and deterministic compatibility assignment with conservative fallback for unresolved techniques. By \emph{headroom} we mean the additional fraction of environment signal that becomes machine-actionable after one such enrichment step is applied to the current corpus. Each step yields auditable intermediate results. The resulting metrics quantify that recoverable margin without claiming full execution fidelity.

\myparagraph{Profile specificity} For each intrusion set $g$, we encode its SUT profile as a binary vector over all software and CVE identifiers in the corpus. Software membership is keyed by \stix{} object reference rather than by name, so alias lists carried on a malware or tool object cannot split that object in the profile. The metric is reproducible over the published bundle, but it is not a product-family canonicalization across distinct STIX software objects. Pairwise Jaccard distances are computed, and the fraction of groups with at least one neighbor within $\delta = 0.1$ is reported as the SUT profile confusion rate. This metric uses unweighted set overlap because the profiles themselves are sparse sets rather than weighted timelines; Jaccard distance is $0$ for identical support and $1$ for disjoint support.

We use $\delta = 0.1$ as a near-neighbor threshold, so two profiles count as confusable only when at least 90\% of their union is shared; equivalently, their Jaccard distance is at most 0.1. Operationally, this is a screening threshold: if two groups differ on fewer than 10\% of the software/CVE support used to build their SUT profiles, a lab built for one should not be treated as actor-specific without additional evidence. Reporting the same rate across $k \geq 1$, $k \geq 2$, and $k \geq 3$ linked-software thresholds separates genuine environmental overlap from ambiguity due to sparse evidence. Section~\ref{sec:analysis} reports the measured values under these rules.

Throughout, the measured object is the public structured CTI surface and the claim-to-evidence boundary it supports; any execution substrate serves only as a compatibility check on that surface, never as the object of measurement.

% ===============================================================
\section{Results}\label{sec:analysis}
We present the results as one evidence stack. First, we ask how far public datasets narrow the candidate environment. Next, we ask which replay substrate those signals justify once backend semantics are made explicit. We then test whether the resulting SUT profiles are sufficiently distinctive for evidence-filtered actor screening, rather than only for coarse lab selection, and finally quantify how much bounded enrichment is recovered without altering the underlying corpus. Unless otherwise stated, all results in this section should be read as lower-bound SUT and backend-family claims derived from structured fields plus the bounded enrichment steps defined in Section~\ref{sec:methodology}.

\myparagraph{Evidence coverage} Of the \enterpriseplatformcount{} active Enterprise techniques, all \enterpriseplatformcount{} (\enterpriseplatformpct\%) carry at least one entry in \texttt{x\_mitre\_platforms}, so platform annotation is effectively mandatory in current ATT\&CK curation. In contrast, \texttt{x\_mitre\_system\_requirements}, the most specific platform field, is populated in \enterprisesystemrequirementscount{} techniques (\enterprisesystemrequirementspct\%). Structured, fine-grained OS and software-version constraints are absent in the Enterprise environment.

Platform-field coverage is also high in Mobile (\mobileplatformpct\%), ICS (\icsplatformpercentage\%), and FiGHT (\fightplatformpercentage\%) under the same active-object filter. CAPEC does not use \texttt{x\_mitre\_platforms} in this dataset because it encodes abstract attack patterns rather than platform-bound behaviors.

\begin{figure}[htpb]
\centering
\makebox[\columnwidth][l]{\hspace*{-0.05\columnwidth}\resizebox{1.05\columnwidth}{!}{\definecolor{acmBlue}{HTML}{4477AA}
\definecolor{acmTeal}{HTML}{228833}
\definecolor{acmSand}{HTML}{EE7733}
\definecolor{acmGrid}{HTML}{D9DDE2}
\begin{tikzpicture}[x=0.50cm,y=0.028cm,font=\small]
  \path[use as bounding box] (-1.40,-33.0) rectangle (17.75,139.0);
  \draw[->] (0,0) -- (17.2,0);
  \draw[->] (0,0) -- (0,118);
  \node[rotate=90,font=\small] at (-1.4,60) {Coverage (\%)};
  \foreach \yy in {0,20,40,60,80,100} {
    \draw[acmGrid] (0,\yy) -- (16.8,\yy);
    \node[anchor=east,font=\small,text=gray,inner xsep=0.8pt,inner ysep=0.2pt] at (-0.14,\yy) {\yy};
  }
  \fill[acmBlue]   (1.0,0) rectangle (1.8,100.0);
  \fill[acmTeal]   (1.9,0) rectangle (2.7,68.2);
  \fill[acmSand]   (2.8,0) rectangle (3.6,2.3);
  \node[above,font=\small] at (1.4,100.0) {100};
  \node[above,font=\small] at (2.45,68.2) {68.2};
  \node[above,font=\small] at (3.2,2.70) {2.3};
  \fill[acmBlue]   (4.0,0) rectangle (4.8,100.0);
  \fill[acmTeal]   (4.9,0) rectangle (5.7,87.1);
  \fill[acmSand]   (5.8,0) rectangle (6.6,0.0);
  \node[above,font=\small] at (4.4,100.0) {100};
  \node[above,font=\small] at (5.45,87.1) {87.1};
  \fill[acmBlue]   (7.0,0) rectangle (7.8,98.8);
  \fill[acmTeal]   (7.9,0) rectangle (8.7,80.7);
  \fill[acmSand]   (8.8,0) rectangle (9.6,1.2);
  \node[above,font=\small] at (7.4,98.8) {98.8};
  \node[above,font=\small] at (8.45,80.7) {80.7};
  \node[above,font=\small] at (9.2,1.70) {1.2};
  \fill[acmBlue!40]   (10.0,0) rectangle (10.8,0.0);
  \fill[acmTeal!40]   (10.9,0) rectangle (11.7,0.0);
  \fill[acmSand!40]   (11.8,0) rectangle (12.6,0.0);
  \node[above,font=\small,text=gray] at (11.3,2) {0};
  \fill[acmBlue]   (13.0,0) rectangle (13.8,100.0);
  \fill[acmTeal]   (13.9,0) rectangle (14.7,63.1);
  \fill[acmSand]   (14.8,0) rectangle (15.6,1.2);
  \node[above,font=\small] at (13.4,100.0) {100};
  \node[above,font=\small] at (14.45,63.1) {63.1};
  \node[above,font=\small] at (15.2,1.70) {1.2};
  \node[font=\footnotesize] at (2.3,-8) {Enterprise};
  \node[font=\footnotesize] at (5.3,-8) {Mobile};
  \node[font=\footnotesize] at (8.3,-8) {ICS};
  \node[font=\footnotesize,text=gray] at (11.3,-8) {CAPEC};
  \node[font=\footnotesize] at (14.3,-8) {FiGHT};
  \node[below,font=\small,text=gray] at (2.3,-12) {$n=691$};
  \node[below,font=\small,text=gray] at (5.3,-12) {$n=124$};
  \node[below,font=\small,text=gray] at (8.3,-12) {$n=83$};
  \node[below,font=\small,text=gray] at (11.3,-12) {$n=615$};
  \node[below,font=\small,text=gray] at (14.3,-12) {$n=566$};
  \node[font=\small,text=black] at (8.3,-33.0) {Corpus};
  \node[anchor=north,fill=white,fill opacity=0.95,text opacity=1,inner sep=2.2pt] at (8.6,136.0) {
    \begin{tabular}{@{}c@{\hspace{0.8em}}c@{\hspace{0.8em}}c@{}}
      \textcolor{acmBlue}{\rule{0.95em}{0.72em}}\ \ $\rho_P$: platform &
      \textcolor{acmTeal}{\rule{0.95em}{0.72em}}\ \ $\rho_S$: software &
      \textcolor{acmSand}{\rule{0.95em}{0.72em}}\ \ $\rho_V$: CVE
    \end{tabular}
  };
\end{tikzpicture}}}
\caption{\label{fig:coverage_cross_corpus}Cross-corpus coverage of structured SUT-relevant signals at attack-pattern level. Platform tags remain broadly available, while software-link and CVE evidence fall sharply relative to platform coverage.}
\Description{Grouped bars comparing platform, software-link, and CVE-link coverage across Enterprise, Mobile, ICS, CAPEC, and FiGHT.}
\end{figure}
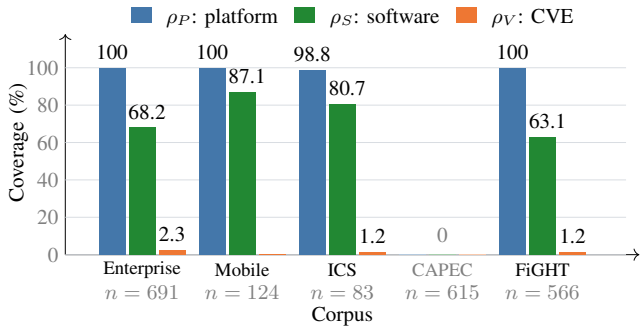

Across datasets, Figure~\ref{fig:coverage_cross_corpus} shows the same pattern: platform-field coverage is high, software-link coverage is moderate, and CVE-linked coverage is low. In CAPEC, the zeros are structural properties of the bundle: active \texttt{attack-pattern} objects link to no \texttt{malware} or \texttt{tool} objects and carry no CVE mentions. They are not missing data caused by the measurement pipeline.

Applying the software-only platform rule to all \enterpriseactivecampaigncount{} Enterprise campaigns yields a platform signal for \enterprisecampaignswithplatformsignalcount{} campaigns (\enterprisecampaignswithplatformsignalpct\%); the remaining \enterprisecampaignsplatformunknowncount{} campaigns (\enterprisecampaignsplatformunknownpct\%) are genuinely unknown under structured fields alone. These campaigns are \emph{FrostyGoop Incident}, \emph{KV Botnet Activity}, \emph{Leviathan Australian Intrusions}, \emph{SPACEHOP Activity}, and \emph{ShadowRay}. Admitting technique-level platform tags would resolve all \enterprisecampaignsplatformunknowncount{} unknowns, but at a cost that confirms why the rule excludes them: single-OS-family campaigns collapse from \campaignswithsingleosfamilycount{} to \platformablationtechniquesingleoscount{}, while triple-OS cross-platform campaigns rise from \campaignswithtripleosfamilycount{} to \platformablationtechniquetripleoscount{} of \enterpriseactivecampaigncount{}. Technique tags remove uncertainty by making almost every campaign appear to target every operating system, which is a compatibility envelope rather than a target, so the software-only rule keeps the \enterprisecampaignsplatformunknownpct\% unknown in exchange for an OS-narrowing signal that targeting claims can use.

Among campaigns with signal, the OS-family distribution (Table~\ref{tab:campaign_os_families}) is inferred from software-linked platform tags in \attack{} rather than from explicit campaign target declarations; therefore, the values should be read as evidence of compatibility. Campaigns may be multi-OS, so OS-family counts can exceed the number of campaigns with signal. When the platform signal is driven by a single cross-platform tool (e.g., Linux/Windows/macOS tags), the inferred set is a compatibility envelope rather than a confirmed target-OS declaration.

\begin{table}[htpb]
\caption{Software-derived OS-family evidence across Enterprise campaigns.}\label{tab:campaign_os_families}
\centering\small
{\rowcolors{2}{stagefill}{white}
\begin{tabularx}{\columnwidth}{@{}Lrrr@{}}
\toprule
\textbf{OS family} & \textbf{Campaigns} & \textbf{\% signal} & \textbf{\% total} \\
\midrule
Windows & \campaignoswindowscount{} & \campaignoswindowspctsignal\% & \campaignoswindowspcttotal\% \\
Linux   & \campaignoslinuxcount{} & \campaignoslinuxpctsignal\% & \campaignoslinuxpcttotal\% \\
macOS   & \campaignosmacoscount{} & \campaignosmacospctsignal\% & \campaignosmacospcttotal\% \\
\midrule
\textit{Platform unknown} & \enterprisecampaignsplatformunknowncount{} & n/a & \enterprisecampaignsplatformunknownpct\% \\
\bottomrule
\end{tabularx}}
\end{table}

How much that signal truly narrows the environment is mixed (Table~\ref{tab:platform_narrowing_quality}). Of the \campaignswithosfamilysignalcount{} campaigns with any OS-family evidence, \campaignswithsingleosfamilycount{} collapse to a single family, \campaignswithdualosfamilycount{} remain dual-family, and \campaignswithtripleosfamilycount{} still expand to a Linux, Windows, and macOS envelope. Within the \onesoftwarecampaignswithossignalcount{} campaigns driven by exactly one linked software item, \onesoftwarecampaignssingleoscount{} narrow cleanly while \onesoftwarecampaignstripleoscount{} still expand to a triple-family compatibility envelope.

\begin{table}[htpb]
\caption{How software-derived platform signal narrows Enterprise campaigns.}\label{tab:platform_narrowing_quality}
\centering\small
{\rowcolors{2}{stagefill}{white}
\begin{tabularx}{\columnwidth}{@{}Lr@{}}
\toprule
\textbf{Case} & \textbf{Campaigns} \\
\midrule
Single OS family & \campaignswithsingleosfamilycount{} \\
Two OS families & \campaignswithdualosfamilycount{} \\
Linux, Windows, and macOS envelope & \campaignswithtripleosfamilycount{} \\
\midrule
Single-OS, one-software case & \onesoftwarecampaignssingleoscount{} \\
Triple-OS, one-software case & \onesoftwarecampaignstripleoscount{} \\
\bottomrule
\end{tabularx}}
\end{table}

Another \campaignswithnonosonlyplatformsignalcount{} campaigns carry only non-OS platform tags, such as network-device labels, and therefore do not narrow the OS family at all under the software-only rule. The platform signal is common, but it often remains broad.

OS-family evidence is, therefore, compatibility evidence for SUT construction, not a standalone attribution signal. The campaigns themselves are not behaviorally sparse: the \enterpriseactivecampaigncount{} Enterprise campaigns use \campaignmeantechniquecount{} techniques on average (median \campaignmediantechniquecount{}) across \campaignmeantacticcoverage{} ATT\&CK tactic categories, and \campaigncompletekillchaincount{} campaigns (\campaigncompletekillchainpct\%) exercise at least one technique in each of the initial-access, execution, discovery, lateral-movement, and exfiltration phases. Only \ieircount{} campaigns (\ieirpct\%) carry no explicit platform or software signal and therefore rely entirely on tactic-level heuristics for environment inference.

To test whether this platform inference is internally stable rather than arbitrary, we compare three signal sources for each campaign: technique metadata, software-linked platforms, and tactic-implied heuristics. They converge for \evidenceconvergenceratepct\% of campaigns. The two divergent cases are \emph{Operation MidnightEclipse} and \emph{RedPenguin}, where technique-level platforms define a broad cross-platform compatibility envelope while software-linked evidence points to Linux and tactic-implied heuristics point to Windows. We keep those conflicts explicit rather than collapsing them into a single OS claim.

Software linkage is common at the campaign and intrusion-set layer. The Enterprise bundle contains \enterpriseactivemalwarecount{} active malware and \enterpriseactivetoolcount{} tool objects (\enterpriseactivesoftwarecount{} total). Of the \enterpriseactivecampaigncount{} campaigns, \enterprisecampaignswithsoftwarecount{} (\enterprisecampaignswithsoftwarepercentage\%) are linked to at least one software object. Among the \enterpriseactiveintrusionsetcount{} active intrusion sets, \enterpriseintrusionsetswithsoftwarecount{} (\enterpriseintrusionsetswithsoftwarepercentage\%) carry at least one software association.

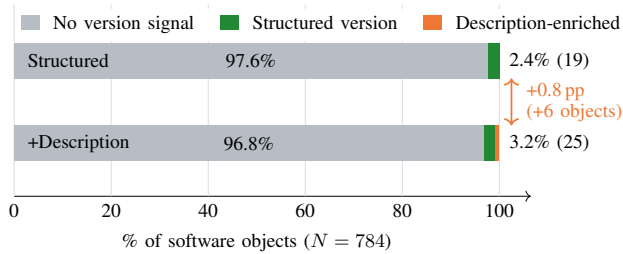
\begin{figure}[htpb]
\centering
\makebox[\columnwidth][l]{\hspace*{-0.05\columnwidth}\resizebox{1.05\columnwidth}{!}{\definecolor{acmGrayFill}{HTML}{B7BDC5}
\definecolor{acmTeal}{HTML}{228833}
\definecolor{acmBlue}{HTML}{4477AA}
\definecolor{acmSand}{HTML}{EE7733}
\definecolor{acmGrid}{HTML}{D9DDE2}
\definecolor{acmGray}{HTML}{8A8F99}

\begin{tikzpicture}[x=0.079cm,y=0.96cm,font=\small]

  \draw[->] (0,0) -- (106.5,0);

  \foreach \x in {0,20,40,60,80,100} {
    \draw[acmGrid] (\x,0) -- (\x,3.25);
    \node[below] at (\x,0) {\x};
  }

  % Row 1: Structured fields only (baseline)
  \fill[acmGrayFill] (0,2.00) rectangle (97.6,2.60);
  \fill[acmTeal] (97.6,2.00) rectangle (100,2.60);

  \node[anchor=west] at (1.2,2.30) {Structured};
  \node at (48.80,2.30) {97.6\%};
  \node[anchor=west,font=\small] at (100.5,2.30) {2.4\% (19)};

  % Row 2: With description enrichment
  \fill[acmGrayFill] (0,0.60) rectangle (96.8,1.20);
  \fill[acmTeal] (96.8,0.60) rectangle (99.2,1.20);
  \fill[acmSand] (99.2,0.60) rectangle (100,1.20);

  \node[anchor=west] at (1.2,0.90) {+Description};
  \node at (48.40,0.90) {96.8\%};
  \node[anchor=west,font=\small] at (100.5,0.90) {3.2\% (25)};

  % Gain annotation
  \draw[<->,acmSand,line width=0.8pt] (102.5,2.00) -- (102.5,1.20);

  \node[
    anchor=west,
    font=\small,
    text=acmSand,
    align=left
  ] at (104.0,1.60)
  {+0.8\,pp\\(+6 objects)};

  % Axis label
  \node[font=\small,align=center] at (50.0,-0.78)
  {\% of software objects ($N = 784$)};

  % Legend
  \node[
    anchor=north west,
    fill=white,
    fill opacity=0.95,
    text opacity=1,
    inner sep=2.2pt
  ] at (0.35,3.18) {
    \begin{tabular}{@{}ccc@{}}
      \textcolor{acmGrayFill}{\rule{0.95em}{0.72em}}\ \ No version signal &
      \textcolor{acmTeal}{\rule{0.95em}{0.72em}}\ \ Structured version &
      \textcolor{acmSand}{\rule{0.95em}{0.72em}}\ \ Description-enriched
    \end{tabular}
  };

\end{tikzpicture}}}
\caption{Enterprise software evidence is rarely version-pinned.}\label{fig:software_specificity}
\Description{Two horizontally stacked bars over active Enterprise software objects, comparing structured-only version evidence with description-enriched evidence. Most objects remain in the no-version segment; a small fraction carry structured version evidence; description enrichment adds a narrow segment; and CPE coverage remains zero.}
\end{figure}

Specificity is the limiting factor. Only \softwarewithversionsignalpercentage\% of the \enterpriseactivesoftwarecount{} software objects contain any version indicator, whether in \texttt{name}, \texttt{aliases}, or \texttt{external\_references}. CPE identifiers, which directly support version-pinned installation, appear in only \softwarewithcpepercentage\% of software objects. Software \emph{presence} is well documented, but software \emph{specificity} for automated SUT construction is largely absent: most software references remain family-level labels without a version pin (Figure~\ref{fig:software_specificity}).

At the campaign level, that bottleneck propagates into profile completeness. Structured fields already support coarse environment profiles for \campaignsuttiercoarsecount{} of the \enterpriseactivecampaigncount{} Enterprise campaigns, which is useful for initial narrowing before analyst refinement, but anchored and exploit-pinned profiles remain rare because they require the version/CPE or campaign-CVE signals that the bundle rarely carries. We next ask whether vulnerability evidence can supply that missing anchor.

Vulnerability evidence is even sparser. CVE identifiers carried by technique, software, campaign, and intrusion-set objects total \cveuniquecount{} unique CVEs. Every extracted identifier passes year-range and sequence-length validation, so format-level extraction precision is total; the only systematic miss is the non-canonical ``CVE~YYYY-NNNN'' spacing variant, which in this corpus occurs only inside free-text \emph{uses} relationship descriptions. Those relationship descriptions are procedure examples, such as a tool exploiting a named CVE, and they annotate how a technique was used rather than which environment a campaign targets. We exclude them from object-level extraction and treat them, like technique-description mentions, as illustrative context rather than as evidence of a structured environment. The bundle contains zero dedicated vulnerability STIX Domain Objects (SDOs), so CVE evidence appears in structured external references or free-text descriptions. Of the \cveuniquecount{} CVEs, \cvetechniqueonlycount{} are illustrative technique-level mentions (for example, \emph{Exploit Public-Facing Application} citing CVE-2016-6662). These are not campaign-specific. The remaining \cveactionablecount{} are actionable because they are linked to software, campaign, or intrusion-set objects.

Coverage remains low at the campaign and group level. Only \entcampaignswithcvecount{} campaigns (\entcampaignswithcvepct\%) carry any CVE reference, and only \entintrusionsetswithcvecount{} intrusion sets (\entintrusionsetswithcvepct\%) have at least one CVE. The gap, therefore, lies in what becomes machine-readable, not just in what is publicly reported: even when evidence of vulnerability exists, it is split between structured external references and free-text descriptions.

\begin{figure}[htpb]
\centering
\makebox[\columnwidth][l]{\hspace*{-0.05\columnwidth}\resizebox{1.05\columnwidth}{!}{\definecolor{acmBlue}{HTML}{4477AA}
\definecolor{acmTeal}{HTML}{228833}
\definecolor{acmSand}{HTML}{EE7733}
\definecolor{acmGray}{HTML}{8A8F99}
\definecolor{acmGrid}{HTML}{D9DDE2}

\begin{tikzpicture}[x=0.253cm,y=0.62cm,font=\small]

  % Axis
  \draw[->] (0,0) -- (30.5,0);

  \foreach \x in {0,5,10,15,20,25} {
    \draw[acmGrid] (\x,0) -- (\x,9.2);
    \node[below] at (\x,0) {\x};
  }

  \node[font=\small] at (15.0,-1.0) {Count};

  % Category labels
  \node[anchor=east] at (-0.4,8.1) {Detected CVEs};
  \node[anchor=east] at (-0.4,6.1) {Actionable CVEs};
  \node[anchor=east] at (-0.4,4.1) {Campaign-linked CVEs};
  \node[anchor=east] at (-0.4,2.1) {Campaigns with $\geq$1 CVE};

  % Bars
  \fill[acmBlue] (0,7.6) rectangle (26,8.6);
  \fill[acmTeal] (0,5.6) rectangle (12,6.6);
  \fill[acmSand] (0,3.6) rectangle (8,4.6);
  \fill[acmGray] (0,1.6) rectangle (5,2.6);

  % Divider
  \draw[densely dashed,acmGrid] (0,3.15) -- (30.8,3.15);

  % Region labels
  \node[font=\small,text=acmGray] at (15.0,3.45)
    {CVE-evidence units};

  \node[font=\small,text=acmGray] at (15.0,1.25)
    {Campaign units};

  % Values
  \node[anchor=west] at (26.5,8.1) {26 (100\%)};
  \node[anchor=west] at (12.5,6.1) {12 (46.2\% of detected)};
  \node[anchor=west] at (8.5,4.1) {8 (66.7\% of actionable)};
  \node[anchor=west] at (5.5,2.1) {5 campaigns (9.6\% of 52)};

\end{tikzpicture}}}
\caption{Operational CVE funnel in Enterprise STIX data. Most detected CVEs never become campaign-level exploit anchors in the Enterprise bundle.}\label{fig:cve_location}
\Description{Horizontal funnel-style bar chart showing attrition from detected CVEs to actionable CVEs, campaign-linked CVEs, and campaigns with at least one CVE signal.}
\end{figure}
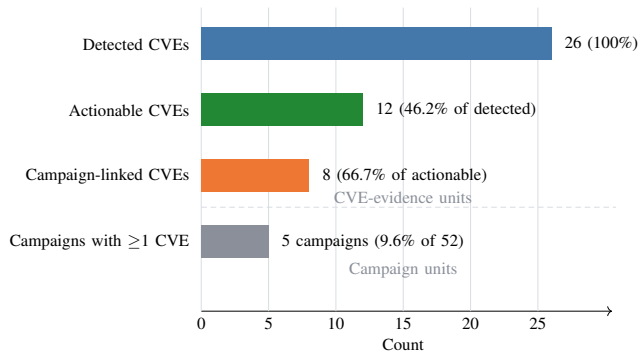

To test how much of this gap is recoverable without changing the corpus, we promote CVE mentions from free-text descriptions into machine-readable CVE slots and then recount coverage. Under this normalization, campaign-level CVE coverage rises from \entcampaignswithcvestructuredcount{}/\enterpriseactivecampaigncount{} (\entcampaignswithcvestructuredpct\%) to \entcampaignswithcvecount{}/\enterpriseactivecampaigncount{} (\entcampaignswithcvepct\%), a gain of \entcampaignswithcveenrichmentgaincount{} campaigns (+\entcampaignswithcveenrichmentgainpp{} percentage points). For intrusion sets, coverage rises from \entintrusionsetswithcvestructuredcount{}/\enterpriseactiveintrusionsetcount{} (\entintrusionsetswithcvestructuredpct\%) to \entintrusionsetswithcvecount{}/\enterpriseactiveintrusionsetcount{} (\entintrusionsetswithcvepct\%), a gain of \entintrusionsetswithcveenrichmentgaincount{} (+\entintrusionsetswithcveenrichmentgainpp{} percentage points). Figure~\ref{fig:cve_location} summarizes the attrition path from all detected CVEs to campaign-level usable evidence.

The gain remains small because most campaigns do not provide explicit evidence of vulnerability. Only \campaignlinkedcvecount{} actionable CVEs are tied directly to campaign objects in the Enterprise STIX dataset, spanning the five campaigns in Table~\ref{tab:campaign_cve_evidence}. Four retain linked software and platform evidence and therefore reach the exploit-pinned tier; \emph{ShadowRay} is the counterexample, carrying CVE-2023-48022 but no linked software object and no software-derived platform signal, so it does not anchor an exploit-pinned lower-bound SUT claim under the conservative rules used in this paper.

\begin{table}[htpb]
\caption{Campaign-linked CVEs in Enterprise ATT\&CK v18.1 and their evidence tier. Entries are from the v18.1 bundle snapshot, extracted on March 5, 2026; CVE year values reflect the assignment year, not the bundle publication year.}\label{tab:campaign_cve_evidence}
\centering\small
{\setlength{\tabcolsep}{2pt}
\rowcolors{2}{stagefill}{white}
\begin{tabularx}{\columnwidth}{@{}>{\raggedright\arraybackslash}p{0.31\columnwidth}>{\raggedright\arraybackslash}p{0.2701\columnwidth}>{\centering\arraybackslash}p{0.06\columnwidth}>{\centering\arraybackslash}p{0.07\columnwidth}L@{}}
\toprule
\textbf{Campaign} & \textbf{CVE(s)} & \textbf{SW} & \textbf{Plat} & \textbf{Tier} \\
\midrule
APT28 Nearest Neighbor Campaign & CVE-2022-38028 & 2 & \checkmark & Exploit-pinned \\
Operation MidnightEclipse & CVE-2024-3400 & 1 & \checkmark & Exploit-pinned \\
Versa Director Zero Day Exploitation & CVE-2024-39717 & 1 & \checkmark & Exploit-pinned \\
SharePoint ToolShell Exploitation & CVE-2025-49704, -49706, -53770, -53771 & 4 & \checkmark & Exploit-pinned \\
ShadowRay & CVE-2023-48022 & 0 & -- & Not anchored \\
\bottomrule
\end{tabularx}}
\end{table}

The contrast between \emph{SharePoint ToolShell Exploitation} and \emph{ShadowRay} makes the boundary concrete. \emph{SharePoint ToolShell Exploitation} reaches the exploit-pinned tier because the bundle carries four campaign-linked CVEs, linked software, and a software-derived platform signal. Even in that strongest case, the structured evidence still stops short of a replay-ready SUT: ATT\&CK names vulnerable software and CVEs, but not the exact product-version tuple, trust boundary, or deployment topology needed to automatically instantiate the lab.

\emph{ShadowRay} fails in the opposite direction. It carries a campaign-level CVE, but without linked software and without a software-derived platform signal, that vulnerability reference cannot be turned into an exploit-pinned lower-bound SUT claim. CVE evidence can only pin down exploitability after software identity and the coarse platform have already been established. Recent automated CVE reproduction pipelines~\cite{ullah2026cvegenie} show that once a CVE is fully resolved to a product and exploit, the operational gap can be closed with high coverage; the bottleneck this paper measures sits one layer upstream, on the CTI side.

The concretization audit provides a bounded downstream probe of the remaining gap. When we intentionally narrow the problem to installable package ecosystems such as \texttt{pip} or \texttt{apt}, only one of the eight campaign/CVE pairs reaches an automatic public path. In the current slice, \emph{ShadowRay} / \texttt{CVE-2023-48022} is the only pair that resolves to an open-package target (\texttt{pip}: \texttt{ray}). The remaining seven pairs still terminate in Windows-component, appliance, or enterprise-server classes that require product-specific reconstruction outside the public path. The automatic path is therefore entirely \texttt{pip}-based; no pair reaches the current \texttt{apt}-materializable path. None of the eight pairs has a direct ATT\&CK target-product binding: the single automatic path comes from curated CVE-plus-package metadata rather than from an ATT\&CK software object that already names the vulnerable product.

Across the corpus, \campaignswithinitialaccesscount{}/\enterpriseactivecampaigncount{} campaigns (\campaignswithinitialaccesspct\%) are linked to at least one Initial Access technique even though CVE evidence remains sparse. Of those, \campaignswithinitialaccessnocvecount{} campaigns (\campaignswithinitialaccessnocvepct\%) lack campaign-level CVE evidence, confirming that Initial Access is often documented without explicit exploit identifiers in the current encoding.

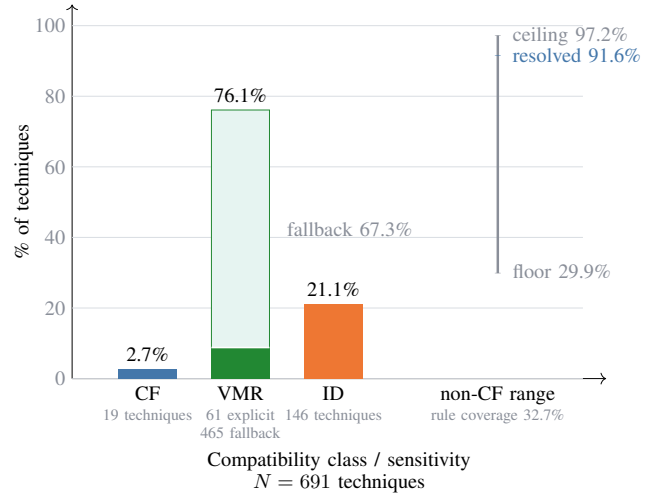
\begin{figure}[htpb]
\centering
\makebox[\columnwidth][l]{\hspace*{-0.05\columnwidth}\resizebox{1.05\columnwidth}{!}{\definecolor{acmBlue}{HTML}{4477AA}
\definecolor{acmTeal}{HTML}{228833}
\definecolor{acmTealLight}{HTML}{E6F4F1}
\definecolor{acmSand}{HTML}{EE7733}
\definecolor{acmGray}{HTML}{8A8F99}
\definecolor{acmGrid}{HTML}{D9DDE2}
\begin{tikzpicture}[x=0.603cm,y=0.055cm,font=\small]
  \path[use as bounding box] (-1.55,-35.5) rectangle (14.6,108.0);
  \draw[->] (0,0) -- (13.8,0);
  \draw[->] (0,0) -- (0,106);
  \node[rotate=90,font=\small] at (-1.25,53) {\% of techniques};
  \foreach \yy in {0,20,40,60,80,100} {
    \draw[acmGrid] (0,\yy) -- (13.2,\yy);
    \node[anchor=east,font=\small,text=acmGray,inner xsep=0.8pt,inner ysep=0.2pt] at (-0.12,\yy) {\yy};
  }
  \fill[acmBlue] (1.2,0) rectangle (2.7,2.7);
  \fill[acmTeal] (3.6,0) rectangle (5.1,8.8);
  \fill[acmTealLight,draw=acmTeal,line width=0.35pt] (3.6,8.8) rectangle (5.1,76.1);
  \draw[white,line width=0.7pt] (3.6,8.8) -- (5.1,8.8);
  \fill[acmSand] (6.0,0) rectangle (7.5,21.1);
  \node[above] at (1.95,2.7) {2.7\%};
  \node[above] at (4.35,76.1) {76.1\%};
  \node[above] at (6.75,21.1) {21.1\%};
  \node[anchor=west,font=\small,text=acmGray] at (5.35,42.45) {fallback 67.3\%};
  \node[below] at (1.95,0) {CF};
  \node[below] at (4.35,0) {VMR};
  \node[below] at (6.75,0) {ID};
  \node[below,font=\scriptsize,text=acmGray] at (1.95,-6.5) {19 techniques};
  \node[below,font=\scriptsize,text=acmGray,align=center] at (4.35,-6.5) {61 explicit\\465 fallback};
  \node[below,font=\scriptsize,text=acmGray] at (6.75,-6.5) {146 techniques};
  \draw[line width=1.1pt,acmGray] (11.0,29.9) -- (11.0,97.2);
  \fill[acmGray] (11.0,29.9) circle (0.085);
  \fill[acmGray] (11.0,97.2) circle (0.085);
  \fill[acmBlue] (11.0,91.6) circle (0.085);
  \node[anchor=west,font=\small,text=acmGray] at (11.18,97.0) {ceiling 97.2\%};
  \node[anchor=west,font=\small,text=acmGray] at (11.18,30.5) {floor 29.9\%};
  \node[anchor=west,font=\small,text=acmBlue] at (11.18,91.6) {resolved 91.6\%};
  \node[below] at (11.0,0) {non-CF range};
  \node[below,font=\scriptsize,text=acmGray] at (11.0,-6.5) {rule coverage 32.7\%};
  \node[font=\small,text=black,align=center] at (6.9,-26.5) {Compatibility class / sensitivity\\$N=691$ techniques};
\end{tikzpicture}}}
\caption{\label{fig:compatibility_distribution}Compatibility-class distribution for active Enterprise techniques. The VMR bar separates explicit VMR evidence from fallback-driven VMR completion; container-feasible coverage remains a small minority under every conservative sensitivity setting.}
\Description{Bar chart of compatibility-class percentages and counts for CF, VMR, and ID across active Enterprise techniques. The VMR bar is split into explicit-rule and fallback-driven portions, and the right side shows a non-CF sensitivity interval with a floor, ceiling, and rule-only-resolved marker.}
\end{figure}

\myparagraph{Backend compatibility} RQ2 asks whether the environment signal measured above is strong enough to support backend claims. Our answer is conservative: CF/VMR/ID is a lower-bound backend taxonomy rather than execution ground truth. Among the \enterpriseplatformcount{} active Enterprise techniques, \compatibilitycontainerfeasiblecount{} (\compatibilitycontainerfeasiblepercentage\%) are container-feasible, \compatibilityvmrequiredcount{} (\compatibilityvmrequiredpercentage\%) are VMR, and \compatibilityinfrastructuredependentcount{} (\compatibilityinfrastructuredependentpercentage\%) are infrastructure-dependent (Figure~\ref{fig:compatibility_distribution}).

Techniques classified as VMR or ID cannot be reproduced faithfully in a container-only environment. Campaigns involving lateral movement, boot-level persistence, or credential access through domain services consistently fall into VMR or ID classes, which sharply limits the coverage of container-based cyber ranges.

The conservative fallback inflates the VMR share, but non-CF techniques remain the majority under every sensitivity setting, spanning \compatibilitynoncffloorpercentage\% to \compatibilitynoncfceilingpercentage\% between the floor and ceiling policies. The interval in Figure~\ref{fig:compatibility_distribution} is therefore a policy-sensitivity band, not a confidence interval: the floor forces unresolved techniques to CF, the ceiling forces them to ID, and the marker reports the explicit-rule subset before fallback completion. All \enterpriseplatformcount{} techniques receive a deterministic class assignment (CF, VMR, ID), supporting automated backend-family assignment even when faithful replay is not supported; the non-CF share remains the majority within every ATT\&CK tactic family, so no single tactic rescues container-only replay on its own. Table~\ref{tab:compatibility_rule_breakdown} details where those assignments come from, confirming that the strongest part of the taxonomy is a lower bound on container-feasible coverage, while the weaker part is conservative VMR 
completion.

\begin{table}[htpb]
\caption{Explicit-rule breakdown for Enterprise compatibility classification.}\label{tab:compatibility_rule_breakdown}
\centering\small
{\setlength{\tabcolsep}{5pt}
\rowcolors{2}{stagefill}{white}
\begin{tabularx}{\columnwidth}{@{}Lcrr@{}}
\toprule
\textbf{Rule evidence} & \textbf{Class} & \textbf{Techniques} & \textbf{Share} \\
\midrule
Domain, identity, or cloud cue & ID & 132 & 19.1\% \\
Lateral-movement tactic fallback & ID & 14 & 2.0\% \\
Kernel or boot pattern & VMR & 34 & 4.9\% \\
Privileged or kernel name pattern & VMR & 27 & 3.9\% \\
Container-compatible targets & CF & 19 & 2.7\% \\
Conservative fallback & VMR & 465 & 67.3\% \\
\bottomrule
\end{tabularx}}
\end{table}

Most ID assignments come from explicit platform cues already present in ATT\&CK, while most VMR assignments come from conservative fallback rather than from a rich machine-readable notion of hypervisor or domain topology. The taxonomy is therefore strongest as a lower bound on container-only coverage: it identifies the small CF minority and the broad non-CF majority, while exact VM layout and multi-host topology remain separate deployment commitments outside the class labels.

To make that reading auditable, we draw both explicit-rule and fallback rows into a class-balanced validation sample: 12 CF, 12 VMR, and 12 ID across 36 rows in total, with 29 explicit-rule rows and 7 fallback rows, so it separately exposes the narrow CF minority, the explicit ID evidence, and the conservative VMR completion. It is a construct-inspection surface: each sampled row can be checked directly against the explicit rule or fallback that produced it.

The compatibility analysis tells us which type of lab a technique requires. The next question is whether those observable SUT properties can distinguish one threat actor from another (RQ3).

\myparagraph{Profile specificity} Of the \enterpriseactiveintrusionsetcount{} active intrusion sets, \sutprofileuniquesoftwarepercentage\% have a unique SUT profile at Jaccard distance $\delta = 0.1$ when profiles use software constraints only. Adding CVE constraints leaves this value unchanged at \sutprofileuniquesoftwarecvepercentage\%. That result is consistent with sparse CVE linkage: only \entintrusionsetswithcvecount{} intrusion sets carry any CVE evidence. The remaining \sutprofileconfusionsoftwarecvepercentage\% of groups still share their SUT profile with at least one other group.

A SUT built for group $g$ can therefore still match profiles of other groups within distance $\delta$, so a successful emulation is not actor-specific by default. The same pattern of ambiguity observed in behavioral attribution studies~\cite{DBLP:conf/uss/SahaMBCVL25} reappears here at the environmental level. Bootstrap resampling with 5{,}000 replicates leaves that reading unchanged: confusion remains within \bootstrapconfusioncilow{}\%--\bootstrapconfusioncihigh{}\%, and the unique-profile rate remains within \bootstrapuniquecilow{}\%--\bootstrapuniquecihigh{}\%. The reading is also flat across the tested near-neighbor thresholds: confusion stays at \deltazerozerofiveconfusionpct{}\% for $\delta \in \{0.05, 0.10, 0.15, 0.20, 0.30\}$.

The ablations point in the same direction. Adding a CVE, platform, or OS family signal does not materially improve separability beyond what software identity alone provides. Compatibility-summary features add 0.6~pp of confusion reduction. A cardinality-preserving null model (\nullmodeliterations{} iterations, randomizing software identities while keeping per-IS counts) places the observed confusion inside its expected interval (\nullmodelconfusionplowpct\%--\nullmodelconfusionphighpct\%, permutation $p=\nullmodelpvalue{}$). The key result is that confusion is largely a low-evidence phenomenon: at $\delta = 0.10$, it disappears once a group carries at least two linked software items.

These rates are reported over different populations, which we make explicit to avoid misreading. The \sutprofileconfusionsoftwarepercentage\% figure is the baseline over \emph{all} active intrusion sets, and it is concentrated in evidence-poor groups: intrusion sets with little or no linked software have near-empty profiles that collide with one another, so most of the baseline confusion does not reflect genuine environmental overlap among well-evidenced actors. Restricting to the subset of groups with at least one linked software item ($k \geq 1$) already lowers confusion to \thresholdkoneconfusionpct\%. Raising the minimum evidence threshold from one linked software item to two then eliminates profile confusion under $\delta = 0.10$: confusion drops from \thresholdkoneconfusionpct\% at $k \geq 1$ to \thresholdktwoconfusionpct\% at $k \geq 2$ and remains collapsed at higher thresholds. These thresholds retain most of the population rather than a sparse tail: of the \enterpriseactiveintrusionsetcount{} active intrusion sets, \thresholdkonesample{} carry at least one linked software item, \thresholdktwosample{} carry at least two, \thresholdkthreesample{} at least three, and \thresholdkfivesample{} at least five, so the $k \geq 2$ subset on which confusion vanishes is the majority of the corpus rather than an effect of small samples. Adding CVE evidence does not materially improve separability at that threshold, so the actor-specific screening claim is strictly a function of software-evidence density. The nearest-neighbor cumulative distributions of software-only and software-plus-CVE profiles are almost indistinguishable across the full distance range, which is the visual counterpart of the unchanged \sutprofileconfusionsoftwarepercentage\% confusion rate above.

\myparagraph{Enrichment headroom} We now quantify the marginal value of the bounded enrichment steps defined above. We report how much additional environment signal becomes machine-actionable when description mining and deterministic compatibility assignment are added on top of structured ATT\&CK fields.

Structured ATT\&CK fields expose version signal for only \softwarewithversionsignalpercentage\% of active software objects. Description-based regex enrichment raises this to \softwareversionenrichedtotalpct\%, a gain of \softwareversionenrichmentgainpp~pp. Campaign-level CVE coverage shows a similarly bounded pattern: structured links alone cover \entcampaignswithcvestructuredpct\% of campaigns, and free-text CVE extraction raises this to \entcampaignswithcvepct\% (+\entcampaignswithcveenrichmentgainpp~pp). Compatibility assignment behaves differently. Explicit rules cover \compatibilityrulecoveragepercentage\% of Enterprise techniques, while a deterministic fallback policy resolves \compatibilitynoncfresolvedpercentage\% of non-CF techniques, yielding a larger \compatibilityresolutiongainpp~pp increase in backend assignment coverage. The largest measured gain, therefore, comes from deterministic compatibility assignment rather than from version or CVE extraction (Table~\ref{tab:bounded_enrichment}).

\begin{table}[htpb]
\caption{Measured headroom from bounded enrichment steps.}
\label{tab:bounded_enrichment}
\centering\small
{\setlength{\tabcolsep}{5pt}
\rowcolors{1}{stagefill}{white}
\begin{tabular}{@{}lrrr@{}}
\toprule
\textbf{Signal} & \textbf{Base} & \textbf{Post-step} & \textbf{Gain} \\
\midrule
Software version signal & \softwarewithversionsignalpercentage\% & \softwareversionenrichedtotalpct\% & +\softwareversionenrichmentgainpp~pp \\
Campaign-level CVE signal & \entcampaignswithcvestructuredpct\% & \entcampaignswithcvepct\% & +\entcampaignswithcveenrichmentgainpp~pp \\
Technique backend assignment & \compatibilityrulecoveragepercentage\% & \compatibilitynoncfresolvedpercentage\% & +\compatibilityresolutiongainpp~pp \\
\bottomrule
\end{tabular}}
\end{table}

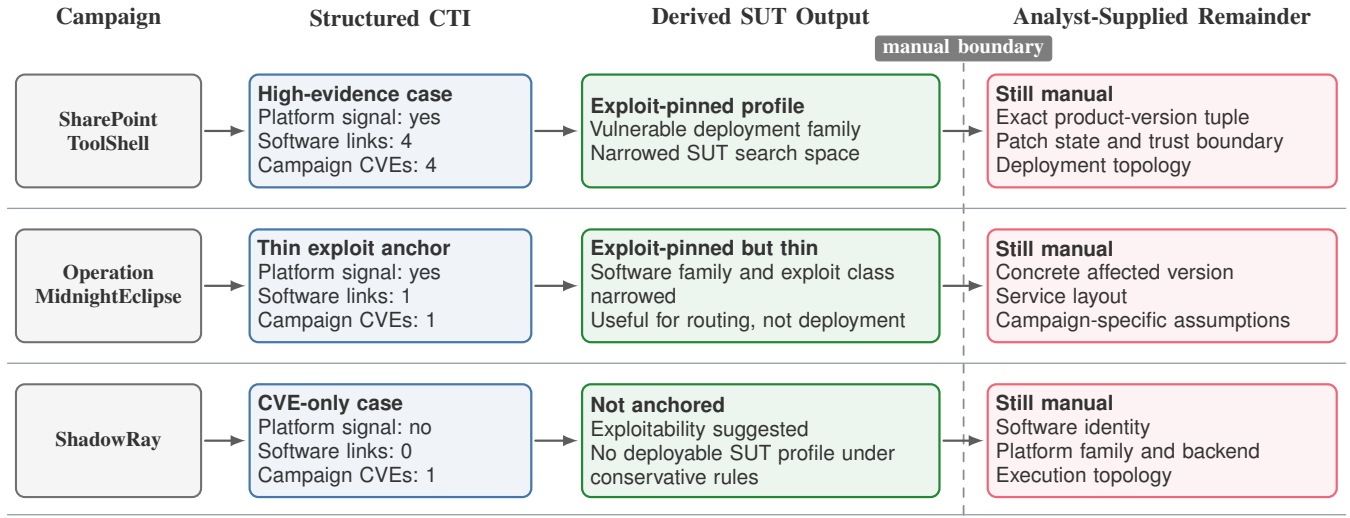
\begin{figure*}[!t]
\centering
\begingroup
\makebox[\textwidth][l]{%
  \hspace*{-0.05\textwidth}%
  \resizebox{1.07\textwidth}{!}{\providecolor{execblue}{HTML}{4477AA}
\providecolor{loopgreen}{HTML}{228833}
\providecolor{ctired}{HTML}{EE6677}
\providecolor{edgegray}{HTML}{333333}
\providecolor{midgray}{HTML}{666666}
\providecolor{dashgray}{HTML}{BBBBBB}
\providecolor{stagefill}{HTML}{F5F5F5}

\begin{tikzpicture}[
  x=1cm,y=1cm,
  font=\sffamily\scriptsize,
  >=latex,
  line cap=round,
  line join=round,
  header/.style={
    font=\bfseries\footnotesize,
    text=edgegray,
    align=center
  },
  campaign/.style={
    draw=edgegray!70,
    rounded corners=3pt,
    fill=stagefill,
    text=edgegray,
    minimum width=2.20cm,
    minimum height=1.34cm,
    align=center,
    font=\bfseries\scriptsize,
    line width=0.7pt
  },
  evidence/.style={
    draw=execblue,
    rounded corners=3pt,
    fill=execblue!8,
    text=edgegray,
    text width=3.15cm,
    minimum height=1.34cm,
    align=left,
    inner sep=2.8pt,
    line width=0.75pt
  },
  derived/.style={
    draw=loopgreen,
    rounded corners=3pt,
    fill=loopgreen!8,
    text=edgegray,
    text width=4.05cm,
    minimum height=1.34cm,
    align=left,
    inner sep=2.8pt,
    line width=0.75pt
  },
  manual/.style={
    draw=ctired,
    rounded corners=3pt,
    fill=ctired!8,
    text=edgegray,
    text width=3.95cm,
    minimum height=1.34cm,
    align=left,
    inner sep=2.8pt,
    line width=0.75pt
  },
  flow/.style={
    ->,
    edgegray!80,
    line width=0.75pt,
    shorten >=1.0pt,
    shorten <=1.0pt
  }
]
  \path[use as bounding box] (-1.50,1.25) rectangle (14.60,7.58);

  % Column centers
  \def\xA{-0.30}
  \def\xB{3.05}
  \def\xC{7.45}
  \def\xD{12.22}

  % Manual boundary centered between derived and manual boxes.
  % Derived right edge = 7.45 + 4.05/2 = 9.475
  % Manual left edge  = 12.22 - 3.95/2 = 10.245
  % Midpoint = 9.86
  \def\xBoundary{9.86}

  % Row centers: slightly more open, still compact
  \def\yA{5.88}
  \def\yB{4.04}
  \def\yC{2.20}

  % Horizontal separators centered between row centers
  \def\lineAB{4.96}
  \def\lineBC{3.12}
  \def\lineBottom{1.32}

  % Headers
  \node[header] at (\xA,7.22) {Campaign};
  \node[header] at (\xB,7.22) {Structured CTI};
  \node[header] at (\xC,7.22) {Derived SUT Output};
  \node[header] at (\xD,7.22) {Analyst-Supplied Remainder};

  % Manual boundary
  \draw[dashed,midgray!75,line width=0.6pt]
    (\xBoundary,1.32) -- (\xBoundary,6.68);

  \node[
    fill=midgray!85,
    draw=midgray!85,
    text=white,
    rounded corners=1.1pt,
    inner xsep=2.5pt,
    inner ysep=1.1pt,
    font=\scriptsize\bfseries
  ] at (\xBoundary,6.86) {manual boundary};

  % Horizontal separators
  \foreach \y in {\lineAB,\lineBC,\lineBottom} {
    \draw[dashgray!95!edgegray,line width=0.45pt]
      (-1.50,\y) -- (14.40,\y);
  }

  % SharePoint ToolShell
  \node[campaign] (c1) at (\xA,\yA)
    {SharePoint\\ToolShell};

  \node[evidence] (e1) at (\xB,\yA)
    {\textbf{High-evidence case}\\Platform signal: yes\\Software links: 4\\Campaign CVEs: 4};

  \node[derived] (d1) at (\xC,\yA)
    {\textbf{Exploit-pinned profile}\\Vulnerable deployment family\\Narrowed SUT search space};

  \node[manual] (m1) at (\xD,\yA)
    {\textbf{Still manual}\\Exact product-version tuple\\Patch state and trust boundary\\Deployment topology};

  \draw[flow] (c1.east) -- (e1.west);
  \draw[flow] (e1.east) -- (d1.west);
  \draw[flow] (d1.east) -- (\xBoundary,\yA) -- (m1.west);

  % Operation MidnightEclipse
  \node[campaign] (c2) at (\xA,\yB)
    {Operation\\MidnightEclipse};

  \node[evidence] (e2) at (\xB,\yB)
    {\textbf{Thin exploit anchor}\\Platform signal: yes\\Software links: 1\\Campaign CVEs: 1};

  \node[derived] (d2) at (\xC,\yB)
    {\textbf{Exploit-pinned but thin}\\Software family and exploit class narrowed\\Useful for routing, not deployment};

  \node[manual] (m2) at (\xD,\yB)
    {\textbf{Still manual}\\Concrete affected version\\Service layout\\Campaign-specific assumptions};

  \draw[flow] (c2.east) -- (e2.west);
  \draw[flow] (e2.east) -- (d2.west);
  \draw[flow] (d2.east) -- (\xBoundary,\yB) -- (m2.west);

  % ShadowRay
  \node[campaign] (c3) at (\xA,\yC)
    {ShadowRay};

  \node[evidence] (e3) at (\xB,\yC)
    {\textbf{CVE-only case}\\Platform signal: no\\Software links: 0\\Campaign CVEs: 1};

  \node[derived] (d3) at (\xC,\yC)
    {\textbf{Not anchored}\\Exploitability suggested\\No deployable SUT profile under conservative rules};

  \node[manual] (m3) at (\xD,\yC)
    {\textbf{Still manual}\\Software identity\\Platform family and backend\\Execution topology};

  \draw[flow] (c3.east) -- (e3.west);
  \draw[flow] (e3.east) -- (d3.west);
  \draw[flow] (d3.east) -- (\xBoundary,\yC) -- (m3.west);

\end{tikzpicture}}%
}
\caption{Worked examples of the SUT-derivation boundary. The three campaigns show where structured CTI ends and analyst-supplied environment details begin: strong exploit-pinned evidence, thin exploit-pinned evidence, and a CVE-only case that remains unanchored.}\label{fig:worked_examples_sut}
\Description{Three-row left-to-right diagram comparing SharePoint ToolShell, Operation MidnightEclipse, and ShadowRay. Each row contains the campaign name, structured CTI evidence, derived SUT output, and analyst-supplied remainder, with a dashed manual boundary separator between the derived output and manual remainder columns.}
\endgroup
\end{figure*}

Taken together, these measurements define the evidence stack available before analyst specification begins (Table~\ref{tab:evidence_stack_sut}). Three representative campaigns in Figure~\ref{fig:worked_examples_sut} show how derivation stops at different missing layers depending on which evidence the corpus carries.

\begin{table*}[!t]
\caption{Measured evidence layers for partial SUT design. Each layer adds a bounded lower-bound claim, not a replay-ready deployment recipe.}\label{tab:evidence_stack_sut}
\centering\small
\renewcommand{\arraystretch}{1.12}
{\rowcolors{2}{stagefill}{white}
\begin{tabularx}{\textwidth}{
>{\raggedright\arraybackslash}p{0.114\textwidth}
>{\raggedright\arraybackslash}p{0.18\textwidth}
>{\raggedright\arraybackslash}p{0.3015\textwidth}
>{\raggedright\arraybackslash}p{0.32\textwidth}}
\toprule
\textbf{Evidence layer}
& \textbf{Measured signal}
& \textbf{What structured CTI already supports}
& \textbf{What remains outside the current corpus} \\
\midrule

Platform coverage
& Near-universal platform tags; sparse system-requirement fields
& Broad OS, device, or cloud-family narrowing and software-derived compatibility envelopes
& Exact OS version, host configuration, deployment placement, and most campaign-specific target declarations \\

Software specificity
& Campaign and intrusion-set software links are common, but version and CPE signal are rare
& Coarse profiles and installable product-family hints; occasional anchored profiles when version/CPE appears
& Product-version tuples precise enough for unattended installation and replay-ready package selection \\

Vulnerability evidence
& Few campaign-linked CVEs; many CVEs remain only in prose
& Exploit pinning for a small minority of campaigns when CVE evidence aligns with software and platform signal
& Machine-readable exploit surface for most campaigns and direct links from vulnerabilities to deployable targets \\

Compatibility classification
& CF/VMR/ID backend classes for every Enterprise technique
& Deterministic backend-family assignment and a lower bound on container-only coverage
& Exact VM count, domain layout, multi-host routing, and service dependencies within the non-CF majority \\

Profile specificity
& Nearest-neighbor confusion over intrusion-set SUT profiles
& A test of whether derived environments are distinctive enough for evidence-filtered actor screening
& Actor-specific interpretation under sparse evidence; environmental overlap remains common until software density rises \\

Bounded enrichment
& Description mining and deterministic fallback assignment
& A reproducible lower bound on recoverable signal without changing the underlying corpus
& Closure of the gap through downstream extraction alone; the measured gains are limited \\

\bottomrule
\end{tabularx}}
\end{table*}

Across these examples, derivation acts as a monotone filter. Platform tags rule out impossible environment families; software links add installable products; campaign-linked CVEs pin exploitability only relative to those products; and compatibility classes choose the minimum backend family. \emph{SharePoint ToolShell Exploitation} is the strongest structured case in the Enterprise corpus and still stops short of replay-ready derivation. \emph{Operation MidnightEclipse} shows exploit pinning on thin evidence, while \emph{ShadowRay} shows that a CVE alone cannot anchor a SUT without supporting software and platform evidence. Analyst specification begins at the first missing layer. A declared step-conditioned overlay applied at execution time remains execution support authored outside the corpus, not new CTI evidence.

Taken together, these measurements define one boundary rather than six separate limitations. Structured CTI supports coarse environment narrowing, deterministic backend-family assignment under conservative rules, and qualified exploit-pinned screening. It does not supply the exact versions, topology, or deployment commitments needed for unattended replay-ready SUT derivation. Actor-specific screening also requires an explicit confusion threshold and enough software density: under this thresholding model, confusion collapses only after an actor carries at least two linked software items. Bounded enrichment recovers the narrow version and vulnerability margins and makes backend commitments explicit, but the analyst-authored remainder remains.

% ===============================================================
\section{Environment Non-Uniqueness}\label{sec:nonuniqueness}

The measurements above show that public structured CTI fixes only part of the executable environment and that emulation authors should separate corpus-supported claims about the environment from analyst-authored lab commitments. This section makes that separation operational and answers RQ4: once the corpus-supported elements are held fixed, does the corpus determine a unique SUT? We answer constructively: it does not.

\myparagraph{Definitions} Let $E$ be the set of materialized SUT elements for a campaign, partitioned by provenance into the \emph{fixed} (corpus-supported) subset $\mathrm{corpus}(E)$ and the \emph{free} subset $\mathrm{free}(E)=E\setminus\mathrm{corpus}(E)$ (analyst-authored plus tool-concretized). Let $\mathrm{fp}(E)$ be an \emph{invariant fingerprint}: the SHA-256 digest of the lexicographically sorted \((\textit{identifier},\textit{value})\) pairs over $\mathrm{corpus}(E)$, with identifiers keyed by \stix{} object reference rather than display name. The sort makes the digest order-independent and ensures that free elements never enter it, so two SUTs share a fingerprint exactly when they agree on every corpus-supported element. Two SUTs $S_1,S_2$ are \emph{campaign-equivalent} when they preserve every corpus-supported element identically ($\mathrm{fp}(S_1)=\mathrm{fp}(S_2)$) and admit execution of the same techniques under the same execution mode. A campaign environment is \emph{non-unique} when two campaign-equivalent SUTs differ in at least one free element. The claim is existential: two such SUTs establish that the campaign is non-unique.

\myparagraph{Construction} A deterministic, seeded generator only alters free elements of a campaign’s standard SUT and checks that fp(·) stays the same. Any change to a corpus-supported element modifies the fingerprint and is not accepted. The resulting provenance tally shows the size of the free region instead of the number of compatible environments. Non-uniqueness is shown constructively. When at least one free element admits multiple versions that preserve all corpus-supported elements and the conditions for executing the technique, the corpus does not define a unique SUT under fingerprint equivalence.

%\Needspace{11\baselineskip}
\myparagraph{Measured free region} The provenance surface makes the construction measurable. Across \provenanceprofilecount{} campaign SUT profiles, \provenancetotalelementcount{} materialized elements are tagged by source: \provenancecorpussupportedcount{} (\provenancecorpussupportedpct\%) are corpus-supported, \provenanceautosutconcretizedcount{} (\provenanceautosutconcretizedpct\%) are AutoSUT concretizations of under-specified corpus signals, and \provenanceanalystauthoredcount{} (\provenanceanalystauthoredpct\%) are analyst-authored. The split is concentrated by dimension. Platform is fully corpus-supported in this materialized slice (\provenanceplatformcorpus{}/\provenanceplatformtotal{}), while topology and staged files are entirely analyst-authored (\provenancetopologyanalyst{}/\provenancetopologytotal{} and \provenanceartifactanalyst{}/\provenanceartifacttotal{}). Software is mostly analyst-authored (\provenancesoftwareanalyst{}/\provenancesoftwaretotal{}); vulnerability evidence is small but more anchored (\provenancevulnerabilitycorpus{}/\provenancevulnerabilitytotal{} corpus-supported). These counts do not enumerate the compatible family; they quantify where the free region lives and explain why a fixed corpus fingerprint can coexist with multiple deployable environments. We deliberately report this determinacy surface rather than a compatible-family cardinality, because a count would depend on the value domain assigned to each free element: a decoy file, a network topology, and a service realization each range over very different spaces, so collapsing them into one scalar would look precise while hiding that modeling choice. The provenance surface instead remains auditable per profile, recording which dimensions the corpus determines, which AutoSUT concretizes from an implied but unspecified need, and which remain of lab authorship.

\myparagraph{Executable witness} The executable witness is a standalone, technique-grounded reference rather than a named ATT\&CK campaign: it instantiates \emph{Exploit Public-Facing Application} (T1190) with the disclosed \emph{CVE-2021-41773}, chosen because it runs end to end and so lets us verify compatibility by real execution rather than by assertion. Its fixed region is the vulnerable Apache~2.4.49 build, its exposed port, and the Linux platform; its free region is a single analyst-authored decoy. Three realizations, the canonical SUT plus two generated variants, share one invariant fingerprint, and each executes the real path-traversal exploit that leaks \texttt{/etc/passwd}, with declared and executed modes identical. The same corpus-supported core, therefore, admits multiple distinct SUTs that all really execute it.

\myparagraph{Coincident witnesses} Two witnesses vary a free service and execute both realizations, across two service classes. The first is a standalone, technique-grounded SSH-pivot reference whose only corpus-supported invariant is the Linux platform of its three hosts. Its canonical realization runs OpenSSH, and a seed-generated variant substitutes Dropbear, a structurally different SSH server. Both run the same set of pivots to completion: network service discovery (T1046), SSH password guessing (T1110.001), lateral SSH access (T1021.004), file discovery (T1083), lateral file staging via SCP (T1570), and remote command execution (T1059.004). The second is an edge-HTTP reference. Its canonical realization runs Apache, and a variant substitutes Nginx, a web server with an independent codebase and configuration model. Both serve a single HTTP Basic Auth realm, so the attacker's service discovery (T1046), password guessing (T1110.001), authenticated access (T1078), and data collection (T1005) run identically against either one. In both witnesses, the substituted element is the service realization itself; the declared and executed modes are identical across all techniques; and the substitution is established by execution rather than by compatibility. Every host runs on an isolated Docker network with no route to the host or the Internet, thereby confining runs to the lab.

%\Needspace{9\baselineskip}
\myparagraph{Structural witness at scale} The named-campaign case is the structural witness. For the \attack{} Enterprise campaign \emph{APT41~DUST}, only the Linux platform is corpus-supported, and the remaining elements are free. Variants substitute \emph{structurally different} services satisfying the same abstract precondition (a remote authenticated service realized as OpenSSH or Dropbear, or a web server as Apache or Nginx) under one unchanged corpus fingerprint. This witness is structural: we report compatibility, not real execution, because the campaign's Linux substrate cannot honestly execute its Windows-oriented techniques. It shows the breadth of the free region rather than an executable consequence.

\myparagraph{Scope} The result is existence, not enumeration: we exhibit at least two campaign-equivalent SUTs per campaign, not the generally unbounded full family. The witnesses span the free region: a corpus that pins enough to be executed leaves a small one, informative service substitutions still execute under moderate ones, and a free region whose realizations need a different substrate is covered only behaviorally. This is a measured property of downstream CTI use: the public ATT\&CK/STIX bundles we evaluate encode adversary behavior and CTI relationships, not complete executable environments.

% ===============================================================
\section{Discussion}\label{sec:discussion}

The results support one operational claim: structured CTI constrains replay environments but does not uniquely determine them.
The measured gaps arise primarily from missing structured fields in public CTI datasets rather than from downstream parsing. The results show a consistent ladder: platform tags and software links often narrow candidate environments; version-anchored software and campaign-linked vulnerability evidence rarely make those profiles replay-ready; and deterministic compatibility assignment mainly improves backend-family inference rather than full reconstruction. Because \compatibilityfallbackassignmentpercentage\% of techniques enter VMR through conservative fallback rather than explicit evidence, the taxonomy is strongest as a lower bound on container-feasible coverage and as a conservative upper bound on how much host or infrastructure support a replay backend may need.

Profile analysis sharpens a second point. A SUT profile that is sufficient for one group is often insufficiently distinctive to justify actor-specific replay claims, which therefore need an explicit profile-confusion threshold. In this corpus, confusion is \thresholdkoneconfusionpct\% at $k \geq 1$ and falls to \thresholdktwoconfusionpct\% at $k \geq 2$ linked software items. Within this screening model, two linked software items are enough to eliminate profile confusion. The same reading survives the $\delta$ sensitivity check.

The missing semantics are concrete rather than abstract. Most campaign-linked software still lacks version information and CPE identifiers; \cvefreetextonlycount{} of \cveuniquecount{} CVEs (\cvefromfreetextpct\%) appear only in natural-language \texttt{description} fields; and the schema carries no field that directly states whether a technique requires a container, a VM, or a multi-host enterprise topology. A CVE identifier alone should therefore not be mistaken for replay-ready exploit evidence, which is consistent with Schloegel et al.'s finding that CVE identifiers are often overread as stronger evidence of real-world impact than the identifier alone supports~\cite{schloegel2025confusing}.

Those same measurements also bound the next system's layer. A deliberately narrowed downstream concretization path from campaign-linked CVEs to installable package ecosystems such as \texttt{apt} or \texttt{pip} currently covers only 1 of 8 campaign/CVE pairs and 1 of 5 CVE-positive campaigns. In the current slice, \emph{ShadowRay} / \texttt{CVE-2023-48022} is the only pair that resolves to an open-package target (\texttt{pip}: \texttt{ray}). The remaining pairs still terminate in Windows-component, appliance, or enterprise-server classes that require product-specific reconstruction outside the public path. The measured coverage collapse shows why this layer is worth separating from the core CTI measurement, and why it cannot replace broader CTI-side environment curation.

Four small schema extensions would measurably reduce specific parts of that gap in the current corpus (Table~\ref{tab:schema_extensions}).

\begin{table}[htpb]
\caption{Schema additions and measured lower-bound deployability impact.}
\label{tab:schema_extensions}
\centering\footnotesize
{\rowcolors{2}{stagefill}{white}
\begin{tabularx}{\columnwidth}{@{}>{\raggedright\arraybackslash}p{0.42\columnwidth}>{\centering\arraybackslash}p{0.25\columnwidth}>{\centering\arraybackslash}p{0.22\columnwidth}@{}}
\toprule
\textbf{Schema addition} & \textbf{Current coverage} & \textbf{Measured gain} \\
\midrule
\path{x_mitre_version_range} & \softwarewithversionsignalpercentage\% software & +\softwareversionenrichmentgainpp~pp* \\
\path{cpe23Uri} on software & \softwarewithcpepercentage\% software & +\softwareversionenrichmentgainpp~pp* \\
\path{x_mitre_sut_class} & \compatibilityrulecoveragepercentage\% rules & +\compatibilityfallbackassignmentpercentage~pp \\
structured CVE relationships & \entcampaignswithcvestructuredpct\% campaigns & +\entcampaignswithcveenrichmentgainpp~pp \\
\bottomrule
\end{tabularx}}

\vspace{1mm}
\footnotesize
*The identical gains arise from the same set of recovered software objects.
\end{table}

These are small additions, not a schema redesign or a replay guarantee. \path{x_mitre_version_range} would represent installable version intervals without requiring exact versions. \path{x_mitre_sut_class} would encode CF/VMR/ID directly and remove heuristic backend assignment. Existing STIX constructs do not already carry that meaning: Course-of-Action objects describe response actions, marking definitions describe handling constraints, and ATT\&CK custom fields such as \path{x_mitre_platforms} expose compatibility or curation metadata rather than a deployability class. Structured CVE relationships would move vulnerability evidence from prose into machine-readable links. The identical +\softwareversionenrichmentgainpp~pp gains for \path{x_mitre_version_range} and \path{cpe23Uri} are expected because the same \softwareversionenrichmentgaincount{} recovered software objects define both margins. Together, these lower bounds give practitioners, CTI curators, and schema designers a concrete design target: shrink the remaining analyst work through version anchoring, explicit vulnerability structure, and declared backend semantics.

This boundary is also orthogonal to the procedural one. A corpus may encode enough behavioral sequence to support replay planning while still lacking the environment evidence needed to instantiate the lab, and the reverse can hold as well. Treating those layers separately prevents replay results from borrowing fidelity from the wrong dimension. A strong replay claim should therefore name both what procedure re-instantiates from the corpus and what environment layer still comes from explicit lab authorship.

That separation gives a practical taxonomy for replay claims. A procedure-complete but environment-incomplete campaign can support a planner benchmark, but its execution result depends on lab choices outside the corpus. An environment-complete but procedure-incomplete campaign can support target reconstruction, but not a faithful campaign replay. When both layers are incomplete, the output is a CTI-inspired scenario rather than a CTI-derived replay. Only when both layers are corpus-supported should a benchmark claim end-to-end replay from structured CTI. AutoSUT measures the environment axis of that matrix, while prior procedural-gap work measures the action-sequence axis; together they define the evidence budget that a replay claim must disclose.

For practitioners, the operational reading should follow the evidence ladder rather than collapsing it into a single, blunt replay claim. Platform and software evidence are already strong enough to eliminate impossible targets and to choose an OS family or backend family early. Campaign-linked CVEs, when they exist, are better read as exploit-screening anchors than as installation-ready replay proof. A corpus-supported profile may therefore justify a claim such as ``VM-backed Windows replay required'' without justifying a specific build, patch level, exposure surface, or network layout. CTI producers can reduce analyst burden by version anchoring software objects and exposing campaign-level vulnerability links as first-class fields. Emulation authors and benchmark designers should label which environment commitments are derived from the corpus and which still come from lab authorship. Emulation engineers can treat backend-family assignment as corpus-supported evidence while keeping topology, trust boundary, and host configuration explicit as analyst-authored layers unless the corpus encodes stronger semantics. Under that reading, reproducibility should stop at the strongest layer the corpus actually supports, so that emulations, benchmarks, and replay claims do not inherit precision the corpus never supplied.

% ===============================================================
\section{Related Work}\label{sec:related}

Jin et al.\ show that CTI sharing provides limited operational value, in part because shared feeds exhibit inconsistent field usage and low structural completeness~\cite{DBLP:conf/ndss/JinKLBKK24}. Prior ATT\&CK surveys identify sparse software versioning as an open problem but do not quantify it~\cite{strom2018mitre,DBLP:conf/uss/VirkudIR00024,furumoto2025ctisurvey}. Broader advanced-persistent-threat (APT) detection surveys likewise emphasize that evaluation depends on representative public datasets and explicit capability criteria~\cite{zhang2024survey}. This line of work establishes that structured CTI is unevenly populated, but its primary object is field completeness rather than environment derivability. The missing question is not whether CTI describes behavior, but whether it encodes enough environment evidence to justify the strongest lower-bound SUT claim before any lab is declared.

Provenance-auditing work surfaces the same dependency at the evaluation layer. Inam et al.\ organize provenance-based system auditing as a capture-to-analysis pipeline and identify reduction and evaluation as persistent bottlenecks~\cite{inam2023sok}. Liu et al.\ show that endpoint-detection conclusions shift with dataset quality and background activity~\cite{liu2025we}, and Bilot et al.\ show that inconsistent evaluation practices and benchmark fragility still hinder practical deployment of provenance-based intrusion detection systems~\cite{bilot2025sometimes}.

Provenance-graph APT detection systems expose a parallel dependency: operational coverage depends on what telemetry and ATT\&CK-linked behaviors the underlying datasets can represent~\cite{milajerdi2019holmes,han2020unicorn,DBLP:conf/uss/Dong0NS0LLX23,DBLP:conf/uss/YangXXLZ23,Li_2024,DBLP:conf/uss/JiaXN0ZW24,DBLP:conf/uss/Zhang0YK0KF25}. Threat-hunting and investigation systems such as \emph{POIROT} and Tactical Provenance Analysis likewise align higher-level attack behavior with audit or endpoint detection and response (EDR) records, but they start from observed telemetry rather than from environment derivation~\cite{milajerdi2019poirot,hassan2020tactical}. Those works measure \emph{behavioral} completeness, not whether structured CTI carries enough environment evidence to instantiate the lab in which emulation is supposed to run.

A separate measurement study examines the procedural version of this question: whether ATT\&CK-in-STIX encodes enough ordering, preconditions, and translation-ready detail for multi-step replay~\cite{ferraz2026proceduralsemanticsgapstructured}. This paper isolates the environment version of the problem: whether the same datasets encode sufficient evidence of platform, software, vulnerability, and deployability to build the SUT in which those steps could run. The two questions require different units of analysis and different failure criteria. Procedural sufficiency asks whether a campaign can be replayed as an ordered sequence; environment sufficiency asks whether any compatible lab can be instantiated at all.

Recent CTI extraction pipelines improve structured recovery of behavioral intelligence from reports, including tactics, techniques, and procedures (TTP) extraction and graph construction~\cite{savat2021extractor,DBLP:journals/corr/abs-2412-10872,cheng2025ctinexus,DBLP:conf/uss/BuchelPLCZBE0GC25}. Attribution-oriented work shows that analyst confidence is often bounded by evidence quality rather than by the absence of behavioral hypotheses, and that high-level indicators may still be too coarse for reliable actor attribution~\cite{DBLP:conf/uss/SahaMBCVL25,DBLP:conf/ccs/YuldoshkhujaevJ25,vanderhorst2025highstakes}. Both lines of work still prioritize \emph{what} happened over \emph{where} execution should occur, so they do not provide a measurement framework for SUT derivation quality.

Dataset building, emulation, and cyber-range systems generally begin after the environment has been designed, rather than deriving it from structured CTI~\cite{cai2026building}. This holds across operator-driven frameworks, automated emulation stacks, ICS testbeds, and recent planning systems guided by large language models~\cite{applebaum2016caldera,redcanary2023art,orbinato2024laccolith,wang2024sands,singer2025incalmo,deng2024pentestgpt,shen2025pentestagent,choi2020expansion,damodaran2025automated}. Effects Language, for example, gives attack graphs executable coordination semantics, but its target operating environment remains external to the formalism; actions query and modify an environment that has already been supplied~\cite{damodaran2025automated}. These systems provide execution logic, planning support, or benchmark labs, but they do not derive deployment-ready SUT specifications from CTI, unlike manually authored, fixed, or benchmarked environments. What is missing is a measure of how much of that environmental burden is already carried by structured CTI and how much falls outside the schema. That missing measurement object is the contribution of this paper.

Along four measurement axes, the distinction sharpens: whether a line of work consumes public structured CTI, measures environment evidence, emits a deployability or backend claim, and evaluates SUT profile specificity. STIX and CTI field-quality studies consume public CTI but only read the field population. Extraction and attribution pipelines consume it and yield partial actor distinctiveness, not SUT specificity. Emulation and range systems sometimes consume public CTI but assume a declared lab rather than measuring environment evidence or emitting a backend claim. This paper reports all four from public ATT\&CK-style structured CTI: it measures the environment evidence already present, issues lower-bound SUT and backend-family claims, and evaluates profile specificity against sparse-evidence conditions. The distinction is narrow but operational. Nearby work measures behavior quality, recovers behavior from reports, or executes behavior on predeclared labs; this paper measures the environment evidence already present in public structured CTI before any lab recipe, range image, or replay substrate is written.

% ===============================================================
\section{Threats to Validity}

This paper measures a lower-bound derivability surface for replay-ready SUTs, not total recoverability from the underlying reports. Construct validity is limited because our measurements capture only what is explicitly encoded in STIX fields. Version and configuration information present in source reports but absent from structured fields is therefore out of scope. The bounded-enrichment results should likewise be read as a recoverable machine-readable signal in the public datasets, not as an upper bound on what deeper report-level reconstruction could recover. Internal validity is also constrained by the asynchronous update pipeline between the ATT\&CK website and downloadable STIX bundles, which may cause extracted counts to underestimate true association rates. This paper measures one ATT\&CK snapshot rather than release-over-release trends, so it does not establish whether environment encoding is improving over time.

External validity is bounded by our focus on public ATT\&CK datasets and by campaign-level platform inference via software-linked platform tags. Commercial CTI feeds and community threat-sharing platforms such as MISP or AlienVault OTX may carry richer software, version, or CPE metadata than the public ATT\&CK-in-\stix{} bundles we evaluate. Multi-platform tools can also inflate OS-family counts without proving the exact operating system targeted in a specific incident. We measure what structured fields encode in public bundles, not what richer report-level or non-public CTI could support.

Profile specificity has the same object-level scope. Object-reference keys avoid name-string alias errors inside one STIX object, but they do not merge distinct software objects through external product-family, CPE, Package URL, or vendor-name normalization. The reported profile-confusion rates are therefore reproducible bundle-level rates, not a claim about every product taxonomy an external enrichment layer could impose.

The compatibility taxonomy introduces a second construct validity risk: some Linux-only techniques may still require richer host semantics than the structured fields reveal. We mitigate that risk by keeping CF narrow and defaulting unresolved cases upward to VMR. The stratified validation sample comprises 36 techniques (12 CF, 12 VMR, and 12 ID), with 29 explicit-rule and 7 fallback rows. We use it as a construct-inspection surface: explicit-rule coverage and fallback sensitivity show how much of the distribution is rule-driven versus how much is driven by conservative completion.

The concretization path is a bounded probe of materialization. Its current public automation narrows the downstream problem to package ecosystems such as \texttt{pip} and \texttt{apt}, so it measures a single materialization slice rather than the full recoverability of product-level exploit targets, as defined by ATT\&CK and external metadata. The same applies to the cross-corpus contrast: Enterprise, Mobile, ICS, CAPEC, and FiGHT are compared using a single extraction pipeline to show what each bundle makes machine-readable, not to claim that their deployment fidelity is otherwise interchangeable.

The lower-bound claims are deliberately asymmetric. Conservative false VMR assignments and conservative ``not anchored'' outcomes prevent overstated deployability and may leave some real replay opportunities uncounted. The rates are lower bounds on corpus-supported derivation, not upper bounds on expert reconstruction from underlying reports. 

The non-uniqueness result (Section~\ref{sec:nonuniqueness}) also serves as a clear example of existence. The CVE witness executes a real exploit, the coincident witnesses execute the same attack across structurally different SSH and HTTP servers, and the named campaign, whose techniques target a different substrate, is shown through structure rather than execution. The contribution is constructive non-uniqueness, which demonstrates that several CTI-consistent environment realizations can exist for the same adversary behavior. We do not aim to list or measure the size of the compatible environment family; we only seek to prove its existence based on the evidence we have.

% ===============================================================
\section{Conclusion}\label{sec:conclusion}

This paper measures the \emph{environment semantics gap} in ATT\&CK-style structured CTI: how far the corpus can support the derivation of replay-ready SUTs before analyst specification resumes. Across five datasets, the evidence forms a clear ladder. Platform annotations and software links often narrow candidate environments, but version anchoring, CPE identifiers, campaign-level CVE links, and backend semantics remain too sparse to justify unattended SUT derivation. Profile specificity improves only after enough software evidence accumulates, and the non-uniqueness witness shows why this boundary matters: structured CTI constrains replay environments but does not uniquely determine them. The design target is therefore concrete rather than rhetorical: add version ranges, structured CVE relationships, and explicit compatibility semantics; until then, emulation authors should separate corpus-supported environment claims from analyst-authored lab commitments so replay statements inherit only the precision the evidence supports.

% ===============================================================
\section*{Ethics considerations}\label{sec:ethics_considerations}

This study analyzes public structured CTI bundles and does not involve human subjects, private operational telemetry, or undisclosed victim data. The main ethical risk is dual use: environment-oriented measurements over CTI could be misread as turnkey replay recipes or used to overstate the extent to which public data already supports emulation by adversaries. We mitigate that risk by keeping the main public path measurement-oriented: it audits the claim surface without synthesizing exploits or executing live intrusion steps.

The optional VM-backed path is likewise bounded. It executes only declared campaign/SUT pairs from the public corpus under local control and does not infer new targets, choose attack steps online, or convert sparse CTI into an autonomous offensive planner. The paper also uses lower-bound language throughout: CVE-linked and compatibility results are reported as bounded-derivation claims, and analyst-authored lab commitments remain explicit whenever the corpus lacks the version, topology, or deployment details needed for replay-ready instantiation.

% ===============================================================
\section*{LLM usage considerations}\label{sec:ai-tools}

The authors used Grammarly and ChatGPT for grammar checking, limited editorial revision, and minor formatting assistance for figures and illustrations. All scientific claims, analyses, experiments, visual representations, and conclusions were produced and verified by the authors.

\bibliographystyle{IEEEtran}
\bibliography{references}


\end{document}